\documentclass[11pt]{article}
\usepackage{a4,epsfig,longtable,supertabular, tabulary}
\usepackage{amsmath,amsthm,amsfonts, mathtools}
\newcolumntype{K}[1]{>{\centering\arraybackslash}p{#1}}
\usepackage{graphicx}
\usepackage{fancyhdr}
\usepackage{color}
\usepackage{bm, bbm} 
\usepackage[kluwer]{harvard}
\usepackage{nicefrac}
\usepackage[ruled,vlined,linesnumbered]{algorithm2e} 
\usepackage{siunitx} 

\numberwithin{equation}{section}

\definecolor{darkblue}{RGB}{24,116,205}

\newtheorem{theorem}{Theorem}[section]
\newtheorem{proposition}[theorem]{Proposition}

\theoremstyle{definition}

\DeclareMathOperator*{\argmin}{arg\,min}

\renewcommand{\P}{\mathbb{P}}
\newcommand{\Q}{\mathbb{Q}}

\newcommand{\EV}[1]{E\left(#1\right)}

\newcommand{\ES}{\mbox{\rm ES}}

\newcommand{\R}{\mathbb{R}}

\newcommand{\F}{\mathcal{F}}
\newcommand{\G}{\mathcal{G}}

\newcommand{\ind}[1]{\mathbbm{1}_{\{#1\}}}

\newcommand{\cds}{\operatorname{cds}}
\newcommand{\ESB}{\operatorname{ESB}}
\newcommand{\EJB}{\operatorname{EJB}}

\begin{document}

\begin{center}
 \large { \Large \bf How Safe are European Safe Bonds? An Analysis from the Perspective of Modern  Credit Risk Models \footnote{%
 We are grateful to Sam Langfield and to two anonymous referees for very useful comments and suggestions.}
 }\\[0.8cm]
{\sc R\"udiger Frey,  Kevin Kurt, Camilla Damian,\footnote{Email: {\tt ruediger.frey@wu.ac.at},  {\tt kevin.kurt@wu.ac.at},  {\tt camilla.damian@wu.ac.at}.}}
  \\[0.5cm]
{\small \it Institute for Statistics and Mathematics, Vienna University of Economics and Business (WU)\footnote{Postal address: Welthandelsplatz 1,  A-1020 Vienna}\\
\today}
\end{center}

\begin{abstract}
Several proposals for the reform of the euro area  advocate the creation of a market in synthetic securities  backed by portfolios of sovereign bonds. Most debated are the so-called European Safe Bonds or ESBies  proposed by \citeasnoun{bib:brunnermeier-et-al-17}. The potential benefits of ESBies and other bond-backed securities hinge on the assertion that these products are really safe. In this paper  we  provide  a comprehensive quantitative  study of  the risks associated with ESBies and related products, using   an  affine credit risk model with regime switching as vehicle for our analysis.
We discuss a recent proposal of Standard and Poors for the rating of ESBies,   we  analyse the impact of model parameters and attachment points on the size and the volatility of the  credit spread of ESBies and we  consider several approaches to assess the market risk of ESBies.  Moreover, we compare ESBies to synthetic securities created by pooling the senior tranche of national bonds as suggested by \citeasnoun{bib:leandro-zettelmeyer-19}.
The paper concludes with a brief discussion of the policy implications from our analysis.
\end{abstract}


\vspace{1cm}

\section{Introduction}

Synthetic securities backed by portfolios of  sovereign bonds  from the euro area  have recently been proposed as a tool  to  improve the stability of the European monetary union and to increase the amount of safe assets in the euro area, see for instance \citeasnoun{bib:EU-17} or \citeasnoun{bib:cepr-on-euro-18}.  The most debated proposal is due to \citeasnoun{bib:brunnermeier-et-al-17}, who advocate the creation of a market in so-called European Safe Bonds or ESBies.  In credit risk terminology, ESBies form the senior tranche of a CDO backed by a diversified portfolio of  sovereign bonds from all members of the euro area. According to \citeasnoun{bib:brunnermeier-et-al-17}, ESBies  would be standardized and issued by tightly regulated private institutions or by a public agency. The junior tranche of the underlying bond portfolio   would be sold in the form of  European Junior Bonds (EJBies) to investors  traditionally bearing default risk, such as hedge funds or insurance companies.

\citeasnoun{bib:brunnermeier-et-al-17} argue that  a liquid market in ESBies would enhance the stability of the euro area in a number of ways: first, it would increase the supply of safe assets in the euro area; second, it would help to break the vicious  circle between bank solvency and the credit quality of sovereigns  created by the fact that most euro area banks hold large amounts of  risky sovereign bonds of the nation state in which they reside; third, it might reduce the distortions on bond markets caused by the flight-to-safety behavior of investors in crisis times.  Moreover, ESBies  respect the \emph{no-bailout clause} and their introduction would not distort \emph{market  discipline}, as the agency issuing ESBies would buy these bonds  at market prices and as sovereigns would remain responsible for their own bonds, which exerts discipline on borrowing decisions.
Another important approach for creating a safe asset for the euro area   consistent with the no-bailout clause is to issue national sovereign bonds in several seniority levels and to pool the bonds from the senior tranche, see for instance~\citeasnoun{bib:leandro-zettelmeyer-19}.
These products and ESBies are  therefore different from eurobonds that are  currently discussed in the context of the Corona crisis. Eurobonds  are jointly issued and guaranteed by all  euro area member states  so that every member state is liable for the entire issuance.  Loosely speaking, ESBies are designed for  improving  the functioning of the euro area ``in normal times'', whereas eurobonds are crisis-intervention instruments.

The potential benefits of ESBies hinge on the assertion that these products are really safe. To address this issue, \citeasnoun{bib:brunnermeier-et-al-17} carry out a simulation study  in an  one-period mixture model where defaults are independent given the aggregate state of the euro area economy.  They find that, with reasonably high  levels of subordination, the  expected loss of ESBies  is comparable to that of triple-A rated bonds. However, their model is calibrated in a fairly ad hoc manner. More importantly, \citeasnoun{bib:brunnermeier-et-al-17} do not study the \emph{market risk} of ESBies (the risk of a change in the market value of these products  due to changes in the credit quality of the underlying bonds or in the  state of the euro area economy).  Now, the  bad performance of many highly rated rated senior CDO tranches during the financial crisis has shown  that the market risk of such products  can  be substantial.  Clearly, a high amount of market risk is inappropriate for a safe asset intended  to serve as collateral in security market transactions, as an investment vehicle for money market funds or as a crisis-resilient  store of value on the balance sheet of banks. A  thorough quantitative analysis of the risks associated with ESBies is thus needed to assess  if these securities  can  in fact perform the function of a safe asset for the euro area. This is the aim  of the present paper.

We propose to work in  a novel  dynamic credit risk model that captures salient features of the credit spread dynamics  of euro area member states and that is at the same time fairly tractable. Such a model is a prerequisite for the analysis of the market risk associated with ESBies.
In mathematical terms, we consider a reduced-form model with conditionally independent  default times; the hazard rate or default intensity of the different obligors is modelled by CIR-type processes whose   mean-reversion level   is a function of  a \emph{common}  finite state Markov chain. Considering a  Markov modulated mean-reversion level permits us to model different regimes, such as  a crisis regime where the default intensity of all  sovereigns is high and an expansionary  regime where all default intensities are low. This  generates  default dependence in a natural way. We successfully calibrate the model to a time series of euro area CDS spreads over the period January~2009 until September~2018. The main  part of the paper is devoted to
the risk analysis  of ESBies and EJBies. We begin by discussing a recent proposal of S\&P for the rating of ESBies, see \citeasnoun{Kraemer2017HowESBies}.   Using novel results on model-independent price bounds for ESBies, we  show that the S\&P proposal is ultra-conservative in the sense that it  attributes  to an ESB the worst rating that is logically consistent with the ratings attributed to the euro area sovereigns. As a next step,  we study the robustness of the  credit spread (or equivalently the risk-neutral expected loss)  of ESBies and EJBies with respect to subordination levels and model parameters. In particular, we consider several parameterizations for the transition intensities  of the  common Markov chain, as these largely drive the default dependence in our model. It turns out that, from this perspective, ESBies are very safe already for low subordination levels (around 15\%), in line with the findings of \citeasnoun{bib:brunnermeier-et-al-17}.

We use several approaches  to gauge the market risk of ESBies. First, we compute spread-trajectories for ESBies via historical simulation, using as input the calibrated trajectories  of the   default intensities and  of the common Markov chain, and we analyse the relation between the attachment point of an ESB and the volatility of the ESB-spreads. Second,  we  carry out a scenario analysis and study how the risk-neutral default probability of these products  is affected by changes in the underlying risk factors.  To robustify our conclusions, we consider also  various {contagion scenarios}. The results of this analysis are more nuanced. For low subordination levels and adverse scenarios (such as the case where the default of a major euro area sovereign leads to a recession in the euro area), the loss probability of ESBies can be fairly large and spread trajectories can be quite volatile. For high subordination levels exceeding 30--35\%, on the other hand, ESBies remain `safe' even in these adverse scenarios. Third,  we  compare the risk characteristics of ESBies to those  of a  safe asset created by pooling the senior tranche of national bonds.
Finally, we use simulations to compute value at risk and expected shortfall for the return distribution of  ESBies.
For this we need to estimate the historical dynamics of the default intensities  and the common Markov chain which is done via  a suitable variant of the EM algorithm.
From this perspective, the market risk of ESBies is fairly low.
Summarizing, we find  that while  in normal times  ESBies are indeed very safe (in fact safer than assets created by pooling the senior tranche of national bonds), they may become  risky  under  extreme circumstances and in contagion scenarios, in particular if the attachment point is not sufficiently high.

We continue with a discussion of the relevant literature. The report of the \citeasnoun{bib:ESRB-18} extends  the quantitative analysis of \citeasnoun{bib:brunnermeier-et-al-17} and considers risk and return characteristics of ESBies and EJBies in  various stress scenarios for default correlation and loss given default; similar issues are studied in  \citeasnoun{bib:barucci-brigo-19} in the context of standard   copula models for defaults. The relevance of market risk for  ESBies is discussed  verbally in \citeasnoun{bib:grauwe-ji-19}. An interesting quantitative analysis of the market associated with ESBies is \citeasnoun{bib:de-sola-perea-et-al-19}. They compute hypothetical spread trajectories for tranches of sovereign bond-backed securities in a copula framework, using  observed bond spreads as input. Techniques from time series analysis (a VAR for VaR analysis and multivariate GARCH modelling)  are used to compute value at risk and marginal expected shortfall for the daily spread change of these tranches.  Further interesting contributions on sovereign bond-backed securities for the euro area are \citeasnoun{bib:langfield-20} or \citeasnoun{bib:cronin-dunne-19}.

Our work is also related to other strands of  the literature on sovereign credit risk, securitization  and  financial innovation.
\citeasnoun{Ang2013SystemicEurope} and \citeasnoun{bib:aitsahalia-et-al-14} carry out interesting empirical work on euro area credit spreads.   \citeasnoun{bib:brigo-et-al-10} give an extensive  discussion of CDO pricing models and their empirical properties before and during the financial crisis, see also \citeasnoun{bib:mcneil-frey-embrechts-15}.   We also use insights from  \citeasnoun{bib:gennaioli-shleifer-vishny-12} or \citeasnoun{bib:golec-perotti-17} regarding  safe assets and financial innovation. Mathematical results on  affine processes  with Markov modulated mean reversion level can be found in \citeasnoun{bib:elliott-siu-09} and in \citeasnoun{bib:vanBeek-et-al-20}.

The remainder  of the paper is structured as follows.
In Section~\ref{sec:model} we formally introduce the model and the relevant credit products.  Section~\ref{sec:calibration} outlines the calibration of  our model to market data. The main part  of the paper is
Section~\ref{sec:risk} where we carry out a thorough analysis of the risks associated with ESBies: in Section~\ref{sec:price-range} we discuss the $S\&P$ proposal for the  rating of ESBies and we relate this to model-independent price bounds,
Section~\ref{sec:staticRisk} focuses on expected loss, while  Sections~\ref{sec:ESB_ts} to \ref{sec:marketRisk} deal with the market risk of ESBies.  In Section~\ref{sec:summary} we summarize the findings from the risk analysis and discuss policy implications.

\section{The Setup} \label{sec:model}

\paragraph{Default model.} Throughout we consider a portfolio of $J$~sovereigns  with default times $\tau^j$ and default indicators \smash{$\ind{\tau^j\le t}$}, $1\leq j\leq J$, defined on a probability space~\smash{$(\Omega,\mathcal{F},\Q)$} with filtration  $\mathbb{G} =(\G_t)_{t \ge 0}$. $\mathbb{G}$ is the global filtration, that is all processes introduced are $\mathbb{G}$ adapted. In financial terms the $\sigma$-field $\G_t$ describes the information available to investors at time $t$.  We assume that $(\Omega,\mathcal{F},\Q)$ supports a $J$-dimensional standard Brownian motion $\bm{W} = (W_t^1, \dots, W_t^J)_{t \ge 0}$ and a finite-state Markov chain $X$, independent of $\bm{W}$, with  state space $S^X = \{1,2,\dots,K\}$ and generator matrix $Q = (q_{kl})_{1 \le k,l \le K}$.  The chain $X$ will be used to model transitions between $K$ different states or regimes of the euro area economy, and for $k\neq l$, $q_{kl}$ gives the intensity of a jump from state $k$ to state $l$.
The measure  $\Q$ is the  risk-neutral measure used for the valuation of ESBies; price  dynamics under the historical measure $\P$ are considered in Section~\ref{sec:marketRisk}.
In the analysis of the model we also use the filtration  $\mathbb{F} = (\F_t)_{t \ge 0}$ that is  generated by the Brownian motion $\bm{W}$ and the Markov chain  $X$.

Our default model under the pricing measure $\Q$ is outlined in the following two assumptions.
\begin{description}
\item[A1)] The default times $\tau^1,\dots, \tau^J$  are conditionally independent doubly stochastic default times with $\mathbb{F}$ adapted hazard rate processes $\gamma^1, \dots ,\gamma^J$, see for instance  \citeasnoun[Chapter 17]{bib:mcneil-frey-embrechts-15}. In mathematical terms, for all $t_1,\dots, t_J >0$  it holds that
    $$\Q\big(\tau^1 \ge t_1, \dots, \tau^J > t_j \mid \F_\infty) = \prod_{j=1}^J \exp \Big ( -\int_0^{t_j} \gamma^j_s \,ds\Big) \,. $$

\item[A2)] The processes $\gamma^1, \dots , \gamma^J$  follow CIR-type dynamics with Markov modulated and time-dependent mean-reversion level, that is
\begin{equation} \label{eq:basic-model}
d \gamma_t^j = \kappa^j(\mu^j(X_t) e^{\omega^j t}  - \gamma_t^j) dt + \sigma^j\sqrt{\gamma_t^j} d W_t^j,\; 1 \le j \le J,
\end{equation}
for constants $\kappa^j, \sigma^j >0, \, \omega_j \ge 0$ and  functions $\mu^j\colon S^X \to (0,\infty)$. For notational convenience, we introduce the vector process $\smash{\bm{\gamma} = (\gamma^1_t, \dots, \gamma^J_t)_{t\ge 0}}$.
\end{description}

\paragraph{Discussion.} For small $\Delta t$ the quantity $ \ind{\tau^j >t } \gamma_t^j \Delta t$ gives the  probability that firm  $j$ defaults in the period $(t,t+\Delta t]$, that is $\gamma^j$ is the \emph{default intensity} of firm $j$.  Assumption~A1 implies that given the path  $(\bm{\gamma}_s(\omega))_{s=0}^\infty $ of the hazard rate process, $\tau^1,\dots,\tau^J$  are independent default times.
Dependence of default events is caused by the special form of the  hazard rate dynamics in A2. More precisely, the assumption  that the mean-reversion levels $\mu^1, \dots,\mu^J$  of the hazard rate processes  depend on  the  {common} finite-state Markov chain $X$ creates co-movement in  the hazard rate of different sovereigns, so that, unconditionally, default times are dependent.  Our setup permits also country-specific fluctuations in hazard rates; these are  generated by the independent Brownian motions $W^1,\dots, W^J$ driving the hazard rate dynamics. Adding the factor $e^{\omega^j t}$ implies that  the mean-reversion level of the hazard rates  is upward-sloping between transitions of $X$. This helps to calibrate the model to the observed term structures of sovereign CDS spreads which are typically upward-sloping as well; see Section~\ref{sec:calibration} for details.

Following \citeasnoun{bib:brunnermeier-et-al-17}, we usually consider  $K = 3$ states of the euro area economy.  In the model calibration in Section~\ref{sec:calibration} we find that, for the vast majority of euro area members, $\mu^j(1) < \mu^j(2)< \mu^j(3)$; that is, the mean reversion level of the hazard rates of euro area members is lowest in state one and highest in state three. This allows us to interpret  these states as expansion (state one), mild recession (state two) and  strong recession (state three).  A statistical analysis  in Section~\ref{sec:marketRisk} shows that  a model of the form \eqref{eq:basic-model} with $K=3$ states  can also be used to describe the evolution of the calibrated hazard rates under the historical measure $\P$.

The default model outlined in Assumptions A1) and A2) is well-suited for a risk  analysis of ESBies. In contrast to the copula models used for instance in \citeasnoun{bib:barucci-brigo-19} or in the work of the  \citeasnoun{bib:ESRB-18},  we  model the dynamic evolution of hazard rates and credit spreads. This  allows us to generate future spread trajectories, which is important in  the analysis of market risk.   By assuming  that the hazard rates depend on the common state of the Euro area economy  we  generate default  dependence in a natural way. This gives a lot of flexibility for the valuation of ESBies.  In fact,  the whole range of arbitrage-free prices of ESBies and EJBies consistent with observed CDS spreads can be obtained within our model if parameters are chosen appropriately; see Section~\ref{sec:price-range} for details. At the same time  the model is fairly tractable: due to the conditional independence assumption it is possible to calibrate the model simultaneously to CDS spreads of all euro area sovereigns\footnote{Without conditional independence, the price of single-name credit derivatives depends  on the default state and the hazard rate of other sovereigns in the portfolio, and  the calibration of the model to single-name CDS spreads is practically possible only for very small portfolios. For instance,  in the Hawkes process model of \citeasnoun{bib:aitsahalia-et-al-14} spillover effects are only studied  for the bivariate case.}
and the form of hazard rate dynamics allows for a fairly efficient computation of  credit derivative prices.

On the other hand, our pricing model with conditionally independent defaults does not allow for \emph{contagion effects} (upward jumps in the credit spreads of non-defaulted sovereigns  in reaction to a default event in the euro area), which  might arise if insufficient measures are taken to mitigate the economic fallout from the default of a major euro area member, see \citeasnoun{bib:cepr-on-euro-18}. This is, however, not an issue for studying the risks associated with ESBies. In fact, with appropriately chosen hazard rate dynamics  our pricing model is able to generate arbitrarily conservative (low) valuations for ESBies. Moreover, contagion matters most in  the analysis  of short term price fluctuations and market risk in Section~\ref{sec:scenarios}, and we  do  consider contagion scenarios  in that context.

\paragraph{Loss process and credit default swaps.} The payoff of credit default swaps (CDSs),  ESBies and EJBies depends on the exact form of the  losses generated by  defaults in the underlying sovereign-debt portfolio. Next we therefore describe  the mathematical  model for the  loss processes that we use in our analysis.
We fix a horizon $T >0$ and a set $\mathbb{T}$ of  payment dates $0 = t_0 < t_1 < \dots < t_N = T$ which, in practical applications, usually correspond to quarterly payments. We define for  $1 \le j \le J$ the cumulative \emph{loss process} $L^j$ of sovereign $j$ by
\begin{equation} \label{eq:loss-process}
 L_t^j = \sum_{n = 1}^N \ind{t_{n-1} < \tau^j \leq t_n} \ind{t \geq t_n} {\delta}^j_{t_n},\quad t \in [0, T]\,,
\end{equation}
where the random variable $\delta^j_{t_n}$ gives the  loss given default (LGD) of sovereign $j$ at time $t_n$.\footnote{We prefer to work with \eqref{eq:loss-process} instead of the more standard definition $L_t^j = \ind{\tau_j \le t} \delta^j(X_{\tau^j})$ as \eqref{eq:loss-process} is more convenient for CDS pricing. In any case,  for $(t_n - t_{n-1})$ small the two definitions of $L^j$ are close to each other.}
 We assume that, given $\F_{t_n}$, the LGD $\delta^j_{t_n}$ is beta distributed with $E\big(\delta^j_{t_n} \mid \F_{t_n}\big) = \delta^j(X_{t_n})$ for a function  $\delta^j : S^X \to (0, 1]$. We further assume that, given $X_{t_n}$, $\delta^j_{t_n}$ is independent of all other model quantities.
Working with a random LGD is realistic and, at the same time, helps to robustify our analysis with respect to the exact values chosen for $\delta^j$.
Given portfolio weights  $w^j >0 $ such that $\sum_{j = 1}^J w^j = 1$, we define the \emph{portfolio loss} by
\begin{equation} \label{eq:portfolio-loss}
L_t =  \sum_{j=1}^J w^j L^j_t\,, \quad t \le T\,.
\end{equation}
 The  cash flow stream of the protection-buyer position in a CDS on sovereign $j$ with spread $x$ and premium payment dates $\mathbb{T}$ can be described in terms of the process $L^j$; it is given by
\begin{equation}
\label{eq:payoffCDS}
    L_t^j - \sum_{t_n \le t} x (t_n - t_{n-1}) \ind{\tau^j > t_n}, \quad  0 \le t \le T.
\end{equation}

\paragraph{ESBies and EJBies.} ESBies have not been issued so far, so there is no description of the payment structure of an actual product and no term sheet. Therefore, we consider stylized versions of these products. These stylized ESBies and EJBies do  capture the essential features of every CDO structure, namely pooling and tranching of default risk, so they suffice to analyze the qualitative properties of ESBies.
Denote by $V_T = 1- L_T$ the normalized value of the asset pool and note that  $V_T = 1$ if there are no defaults in the portfolio. The constant $\kappa \in (0,1) $ denotes the lower (upper) attachment point of the senior (junior) tranche. Then the payoff of a stylized ESB respectively a stylized  EJB at $T$ is defined to be
\begin{align}\label{eq:ESB-payoff}
\ESB_T &= \min( V_T , 1-\kappa) = V_T - (V_T - (1-\kappa))^+ = (1-L_T)- (\kappa - L_T)^+ \,,\\
\label{eq:EJB-payoff}
\EJB_T &= (V_T - (1-\kappa))^+ = (\kappa - L_T)^+\,.
\end{align}
In this way, the EJB bears the first $100 \kappa$ percent of the loss in the portfolio, if the loss exceeds $\kappa$, the ESB is affected as well. While stylized ESBies and EJBies are path independent, in the sense that their payoff is a function of the portfolio loss at the maturity date $T$ only, our analysis is easily extended  to path dependent payoffs.

Note that, by definition, we have the following put-call-parity-type relation for the payoff of a stylized ESB and a stylized EJB with identical attachment point $\kappa$
\begin{equation} \label{eq:put-call-parity}
\ESB_T + \EJB_T = V_T \text{ and hence } \ESB_T = (1-L_T) - \EJB_T\,.
\end{equation}

\paragraph{Pricing.}  For simplicity, we assume that the risk-free short rate is constant and equal to $r \ge 0$. We introduce the money market account  $B_{t, s} = \exp(r(s - t))$, $s>t$, so that $B_{t,s}^{-1}$ is the discount factor at time $t$ for a payoff due at time $s$.  We use standard risk-neutral valuation for the pricing of credit derivatives. Hence  the price at $t$ of any integrable $\G_s$ measurable contingent claim $H$ is  equal to
$H_t = \EV{B(t,s)^{-1} H \mid \G_t}$, where the expectation is taken under the risk-neutral measure $\Q$.

For further use we introduce some notation related to the pricing of ESBies.
Let $\bm{L}_t = (L_t^1,\dots, L_t^J)$. The price of an ESB at time $t \in \{t_0, t_1, \dots, t_N\}$ is given by
\begin{equation}\label{eq:def-ESB-price}
\EV{ B_{t,T}^{-1} ((1-L_T)- (\kappa - L_T)^+)  \mid \G_t } =: h^{\ESB,\kappa}\left(t,X_t,\bm{\gamma}_t, \bm{L}_t \right)
\end{equation}
for a suitable function $ h^{\ESB,\kappa}$. This follows from the fact that the processes $\big ( X_{t_n}, \bm{\gamma}_{t_n}, \bm{L}_{t_n})_{n = 0}^N$ are jointly Markov; we omit the details.
Similarly, the price of an EJB is given by
\begin{equation}\label{eq:def-EJB-price}
h^{\EJB,\kappa}\left(t,X_t,\gamma_t, \bm{L}_t\right) :=\EV{B_{t,T}^{-1} (\kappa - L_T)^+)  \mid \G_t}.
\end{equation}
The key tool for the numerical computation of derivative prices is the extended Laplace transform of the hazard rates. For Markov
modulated CIR processes this transform is available in almost closed form; see Appendix~\ref{app:pricing} for details.

\section{Calibration}\label{sec:calibration}

\paragraph{Data and calibration design.}
The available data consist of  weekly CDS spread quotes  from ICE data services for ten euro area sovereigns and  times-to-maturity equal to 1, 2, 3, 4 and 5 years over the period January 7, 2009 until September 3, 2018, giving rise to 510 observation dates.
The sovereigns used in our analysis are Austria (AUT), Belgium (BEL), Germany (DEU), Spain (ESP), Finland (FIN), France (FRA), the Republic of Ireland (IRL), Italy (ITA), the Netherlands (NLD) and Portugal (PRT), making up more than 90\% of the euro area  GDP in 2018.
Table~\ref{tab:summaryCDS} reports  summary statistics (sample mean, sample standard deviation, minimum and maximum)  of the CDS spreads, together with  the most recent Standard \& Poor's credit-rating of the ten sovereigns.
Average spreads vary considerably across countries and, with the exception of Ireland, the term structures of the average spreads is  upward sloping.

We calibrate  the model by minimizing the sum of squared differences between the CDS spreads observed on the market  and the spreads generated by the  model.
In order to reduce the dimension of the parameter space, we fix the mean function  and the concentration parameter  of the  beta distribution of the loss  given default $\delta_{t_n}^j$  at the outset.\footnote{The parametrization in terms of mean and concentration parameter is a useful alternative to the standard representation of the beta distribution. Denote by  $g(x;a,b) = \beta(a,b) x^{a-1}(1-x)^{b-1} \ind{x \in [0,1]}$ the beta density for given parameters $a,b>0$. Then the mean is given by $a/(a+b)$ and the concentration parameter is $\nu:=a+b$.  A high value of $\nu$  implies that the LGD is very concentrated around its conditional mean.}
The distinct values for the mean function ${\delta}^{j}(\cdot) $   can be found in Table~\ref{tab:lgd} in the appendix.
Following  \citeasnoun{bib:brunnermeier-et-al-17} we assume that the mean LGD is highest in state~3 (the recession state) and lowest in state 1. Moreover, we work with a concentration parameter $\nu=1.5$; this is a conservative  choice as it leads to a fairly widespread  LGD distribution (and we will see in Section~\ref{sec:risk} that a widespread LGD distribution makes ESBies riskier).  While the order of magnitude of the mean LGD is in line  with the  sovereign-debt literature, the exact numerical  values for the mean  LGD and the concentration parameter $\nu$ were handpicked by the authors.\footnote{In fact, $\nu$  cannot be calibrated from CDS spreads, since  model CDS spreads  depend only on the conditional mean of the LGD.}
In Section~\ref{subsec:tranche-and-pool} we therefore study the robustness of our risk analysis   with respect to the parameters of the LGD distribution.

In the  calibration we work  with $K=3$ states of $X$ and  we use the EONIA at date $t$ as a proxy for $r_t$.
We have to determine the trajectories of $X$ and $\bm{\gamma}$ and the parameters $(\Theta^j ,  \sigma^j,Q)$ with $\Theta^j = (\mu^j(1), \mu^j(2), \mu^j(3), \kappa^j,\omega^j)$. We impose the restriction that all parameters are nonnegative, and, to preserve the interpretation of $\mu^j(\cdot)$  as mean-reversion level,  we impose the uniform lower bound $\kappa^j >0.1$ for all $j$.
We use $s_0 < s_1 < \dots < s_M$ to denote the observation dates and  we write $\{\bm{\gamma}_{s_m}\} = \{\bm{\gamma}_{s_0}, \dots, \bm{\gamma}_{s_M}\}$ and $\{X_{s_m}\} = \{X_{s_0}, \dots, X_{s_M}\}$ for trajectories of $\bm{\gamma}$ and $X$. Denote by $\cds_{s_m, u}^j$ the market CDS spread  with  time to maturity $u$  at time $s_m$ and by $\widehat{\cds}(u,\gamma^j_{s_m}, \Theta^j, \sigma^j, Q,X_{s_m})$ the corresponding model spread as function of $\gamma^j_{s_m}$, $X_{s_m}$ and of the model parameters.
We  determine the  model parameters and the realized trajectories $\{\bm{\gamma}_{s_m}\}$ and $\{X_{s_m}\}$ by minimizing the global calibration error
$$
\sum_{m=1}^{M} \sum_{j=1}^{J} \left(\cds_{s_m, u}^j - \: \widehat{\cds}(u,\gamma^j_{s_m}, \Theta^j, \sigma^j, Q,X_{s_m}) \right)^2\,,
$$
using a set of modern optimization algorithms. For this we use an iterative approach which is described in detail in Appendix~\ref{app:calibration}.

\paragraph{Results.}
We implement the calibration methodology on the full time series of available CDS data. We use  maturities of one  and five years since  one-year CDS spreads are particularly informative regarding the current value of the hazard rates whereas five-year CDS markets  are most liquid.
To assess the quality of the calibration, we report in Table \ref{tab:pricingErr} the  \emph{root mean squared error} (RMSE) for all countries and both maturities. As RMSE is scale-dependent,  we also report a relative measure for the calibration error, namely   the  \emph{mean  absolute percentage error} (MAPE). The quality of the calibration is illustrated further in Figure~\ref{fig:CDSseries} in Appendix~\ref{app:calibration}, where we plot the  time series of CDS spreads together with the model prices and the absolute pricing errors for the Germany and Italy.
Given  the complexity of the calibration task, we conclude that the calibrated model fits the observed CDS spreads reasonably well. 


\begin{table}[ht]
\centering
\resizebox{\columnwidth}{!}{%
\begin{tabular}{@{\extracolsep{5pt}}lSSSSSSSSSS}

 {Mat.} & {AUT} & {BEL} & {DEU} & {ESP} & {FIN} & {FRA} & {IRL} & {ITA} & {NLD} & {PRT} \\
  \hline
& \multicolumn{10}{c}{\underline{RMSE (bp)}} \\
1 & 6.36 & 8.70 & 4.73 & 39.72 & 3.58 & 6.75 & 4.58 & 0.45 & 2.90 & 1.49 \\
5 & 15.58 & 15.30 & 9.26 & 45.41 & 6.73 & 13.07 & 40.10 & 34.76 & 10.60 & 66.56 \\
& \multicolumn{10}{c}{\underline{MAPE ($\%$)}} \\
1 & 27.25 & 27.48 & 20.55 & 34.47 & 29.60 & 23.94 & 5.27 & 0.45 & 6.17 & 1.86 \\
5 & 26.29 & 24.36 & 20.85 & 20.11 & 15.79 & 21.62 & 28.54 & 15.38 & 20.74 & 27.59 \\
   \hline
\end{tabular}
}
\caption{\small \label{tab:pricingErr} Calibration error in basis points for maturities of one and five years.}
\end{table}

Tables~\ref{tab:paramValMu} and \ref{tab:paramValQ} report the parameter values resulting from the calibration.
First, note  that $\mu^j(1) < \mu^j(2) < \mu^j(3)$ for all sovereigns except Germany, where  $\mu(1) \ge \mu(2)$.\footnote{This reverse ordering is easily explained by  Germany's prominent role as the euro area's safe haven in times of financial distress.} The uniform ordering of the mean-reversion levels allows us to  interpret  the states of $X$ as expansion, mild and strong recession, and it provides clear evidence that there is strong co-movement in the market's perception of the credit quality of euro area members.
The resulting ordering of the mean reversion levels is also in line with the empirical findings of \citeasnoun{bib:altman-05}, who show that the LGD of corporate bonds is typically positively correlated with their respective default rates (recall that the mean function of the LGD satisfies $\delta^j(1) \leq \delta^j(2) \leq \delta^j(3)$ for all sovereigns).
The mean reversion speed $\kappa^j$ is quite low for all countries, and for four of them (Austria, Belgium, Finland and France) it is equal to the exogenously imposed lower bound of $0.1$.
Consequently, market participants expect idiosyncratic credit shocks to have a long-lasting effect across the term structure of CDS spreads.
The motivation for including the parameter $\omega^j$ is to better capture the upward sloping term structure of most of the CDS series.
In fact,  for $\omega^j=0$ and unrestricted $\kappa^j$, the calibration frequently leads to  negative   values for  $\kappa^j$ --- a common phenomenon also reported e.g.\ in \citeasnoun{Ang2013SystemicEurope}.
Table~\ref{tab:paramValQ} reports the estimate of the generator matrix $Q$. Note that, for the estimated $Q$, transitions to non-neighbouring states have zero probability.

\begin{figure}
\centering
\includegraphics[width=0.8\textwidth]{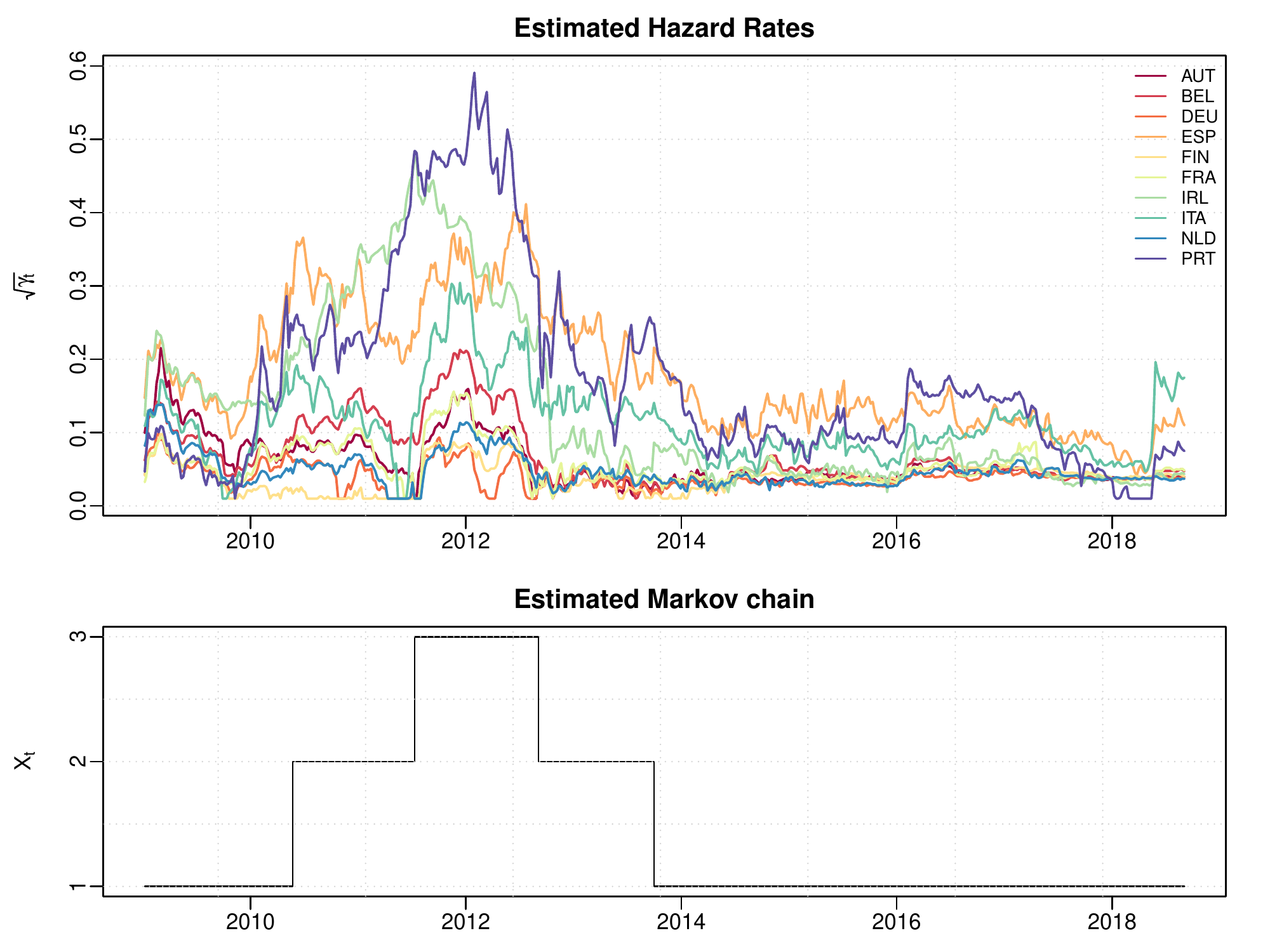}
    \caption{\small Time series plots of the estimated hazard rates and the calibrated Markov chain. Note that  we graph $\sqrt{\gamma_t}$  as this is the natural scale for a CIR process.}
    \label{fig:gammaX}
\end{figure}

Figure~\ref{fig:gammaX} plots the calibrated hazard rates together  with the calibrated trajectory of the Markov chain $X$.
The process $X$ remains in state one for most of the sample period, the only exceptions occur at the height  of the European sovereign debt crisis from mid-2010 until late 2013, when the chain visits states two and three before settling in state one again.
In general, the paths of the hazard rates are in line with the movement of the Markov chain; exceptional individual events such as the rise of the Portuguese hazard rates  at the beginning of 2016 or the sudden upward movement of Italian rates during mid-2018 are of idiosyncratic nature.

\begin{table}[h!]
\renewcommand*{\arraystretch}{1.1}
\centering
\resizebox{\columnwidth}{!}{%
\begin{tabular}{@{\extracolsep{5pt}}lSSSSSSSSSS}
 {Param.} & {AUT} & {BEL} & {DEU} & {ESP} & {FIN} & {FRA} & {IRL} & {ITA} & {NLD} & {PRT} \\
  \hline
$\mu(1)$ & 0.0049 & 0.0044 & 0.0027 & 0.0053 & 0.0047 & 0.0051 & 0.0177 & 0.0710 & 0.0045 & 0.0656 \\
  $\mu(2)$ & 0.0049 & 0.0128 & 0.0001 & 0.0189 & 0.0048 & 0.0085 & 0.0746 & 0.0727 & 0.0049 & 0.2030 \\
  $\mu(3)$ & 0.0424 & 0.0486 & 0.0338 & 0.0558 & 0.0209 & 0.0502 & 0.2498 & 0.4099 & 0.0458 & 0.2115 \\
  \hline
  $\kappa$ & 0.1000 & 0.1000 & 0.1076 & 2.6879 & 0.1000 & 0.1000 & 0.1920 & 0.1215 & 0.1197 & 0.1181 \\
  $\omega$ & 0.1730 & 0.1924 & 0.1534 & 0.0826 & 0.1592 & 0.1916 & 0.0004 & 0.0011 & 0.0994 & 0.0219 \\
  $\sigma$ &  0.1447 & 0.1152 & 0.0872 & 0.2472 & 0.0639 & 0.1075 & 0.1994 & 0.2113 & 0.0925 & 0.3162 \\
   \hline
\end{tabular}
}
\caption{\small  \label{tab:paramValMu} Calibration results:   parameters of hazard rate dynamics .}
\end{table}

\begin{table}[h]
\centering
\begin{tabular}{ lSSS}
                            &{State 1} & {State 2} & {State 3} \\
                            \hline
   {State 1 (expansion)} & -0.1421 & 0.1421 & 0.0000 \\
   {State 2 (mild recession)} & 0.5843 & -1.1685 & 0.5843 \\
   {State 3 (strong recession)} & 0.0000 & 0.9630 & -0.9630 \\
   \hline
\end{tabular}
\caption{\small \label{tab:paramValQ} Calibration results:  generator matrix $Q$ of $X$.}
\end{table}

\section{Risk analysis}\label{sec:risk}

After the successful calibration of  our model, we may now analyze the risks associated with  ESBies. We begin with a short overview. In Section~\ref{sec:price-range} we discuss a recent proposal of the rating agency Standard and Poors (S\&P) for the rating of ESBies (\citeasnoun{Kraemer2017HowESBies}) and we relate the S\&P proposal to a worst-case default scenario where the arbitrage-free price of  ESBies attains its lower bound.
In Section~\ref{sec:staticRisk} we compute the risk-neutral expected loss (or equivalently the credit spread) of ESBies as a function of the attachment point $\kappa$ for different parameter sets. We consider  a {base} parameter set  corresponding largely  to the parameters obtained in the model calibration of Section~\ref{sec:calibration},  two {crisis sets} with higher default correlation and an extremal distribution that corresponds  to the worst-case default scenario.

The  subprime credit crisis has shown that the  expected loss at maturity  gives only limited information regarding  the riskiness of tranched credit products such as  ESBies. In fact, the market value of   AAA-rated senior tranches of mortgage backed securities (MBS) fell sharply during the crisis (some were even downgraded), creating huge  losses for many  MBS investors. To analyze if ESBies can perform all functions of a safe asset, we thus need to take a closer look at the associated market risk. We do this in several ways. First,   we use a historical  simulation approach and compute  credit spread trajectories of ESBies for different attachment points, using as input the calibrated  trajectories $\{X_{s_m}\}$ and $\{\bm{\gamma}_{s_m}\}$  from Section~\ref{sec:calibration}.  This analysis gives useful information on  the relation between $\kappa$ and the volatility of ESB credit spreads. Second, many potential ESB investors, such as managers of money market funds,  are  extremely risk averse so that  ``behavior in (quasi) safe asset markets may  be subject to sudden runs when new information suggests even a minimal chance of a loss'' \cite{bib:golec-perotti-17}.  In Section~\ref{sec:scenarios} we therefore study how the risk-neutral \emph{loss probability} $\Q(L_T >\kappa)$ of ESBies is affected by changes in the underlying risk factors.
To guard against  model  misspecification and to incorporate stylized facts regarding investor behavior on markets for safe assets, we  include  various {contagion} scenarios into this  analysis.
Third, we  compare the risk profile  of ESBies to that  of a  safe asset created by pooling the senior tranche of national bonds and we study the robustness of both product classes with respect to  the LGD distribution, see   Section~\ref{subsec:tranche-and-pool}. In Section~\ref{sec:marketRisk} we finally  use simulations to study Value at Risk and Expected Shortfall for the mark-to-market loss of ESBies. For this we resort to the   model dynamics under the real-world measure.

\subsection{The weak-link approach of S\&P  and worst-case default scenarios} \label{sec:price-range}

 In a recent technical document, \citeasnoun{Kraemer2017HowESBies} discusses how the rating agency Standard and Poors (S\&P)
 would determine a rating for  ESBies and EJBies. The proposed methodology is termed \emph{weak-link approach}. The S\&P proposal  has led to a lot of discussion since it  associates a  BBB- rating to an ESB with attachment point $\kappa = 30\%$ (given sovereign-bond ratings of 2017), which is at odds with the idea that ESBies are safe assets meriting top ratings.

To facilitate the description of the approach,  we  assume that the sovereigns are ordered according to their rating, so that  sovereign one has the best rating and sovereign $J$ has the worst  rating.  Given  an ESB with attachment point $\kappa$, define the index $j^*$ by
$j^* = \max \{1 \le j \le J \colon \sum_{i=j}^{J} w^i \ge \kappa \} .$
Then, under the weak-link approach,  the  ESB is assigned the rating of  sovereign $j^*$.
The  assumption underlying  this approach is that ``sovereigns will default in the order of their ratings, with lowest rated sovereigns defaulting first" \cite[Page 4]{Kraemer2017HowESBies} and that the LGD of all sovereigns is equal to one,  so that
the ESB  incurs a loss as soon as the sovereign $j^*$ defaults.

In this section we show that  the weak-link  approach  is extremely conservative in various respects.
We begin by a concise mathematical description. We drop the time index and consider sovereign debt portfolios with generic loss variables $L^j = \delta^j \ind{\tau^j \le T} $,  $1 \le j \le J$, with values in the interval $[0,1]$.
We assume that the expected loss  of each sovereign is fixed, that is $\EV{L^j} = \bar \ell^j$ for a constant $\bar \ell^j  \in [0,1]$. This is a stylized way of calibrating the model to given  ratings or credit spreads of the sovereigns. We order the sovereigns  by their credit quality and assume that  $\bar \ell^1 \le \bar \ell^2  \dots\le \bar \ell^J $.   Next, we define loss variables that represent the default  scenario of the weak link approach. Fix some standard uniform  random variable $U$ and  define  the loss vector $\bm{L}^* = (L^{*}_1, \dots, L^*_J)$ by
\begin{equation}\label{eq:def-L^star}
L^*_j = \ind{U  > 1-  \bar \ell^j }\,, \quad 1 \le j \le J
\end{equation}
Clearly, $E\big (L^*_j\big)= \Q(L^*_j =1)\Q( U  > 1-  \bar \ell^j) = \bar \ell^j$, so that $\bm{L}^*$ respects the expected-loss constraint. Moreover,  under~\eqref{eq:def-L^star}   sovereigns default exactly in the order of their credit quality with sovereign $J$ defaulting first and  $\delta^j = 1$ for all $j$, that is $\bm{L}^*$  is indeed a mathematical model for  the weak link approach. Note that  the loss vector  $ \bm{L}^* $ is  \emph{comonotonic} since  its components are given by increasing functions of the same one-dimensional random variable $U$, see \citeasnoun[Chapter~7]{bib:mcneil-frey-embrechts-15}. Hence, under the weak-link approach, diversification effects between euro area members are ignored completely.

The next result shows that the loss variables in  \eqref{eq:def-L^star} can be interpreted as  worst-case default scenario in the sense that they minimize the value of ESBies over all loss variables that respect the  expected loss constraints. Hence, the price of an ESB under the worst-case scenario is a lower bound for the arbitrage-free price of that bond in any model consistent with these constraints. In the rating context this means  that
the weak link approach associates with an  ESB the worst rating  logically consistent with the ratings of the individual  euro area sovereigns.
\begin{proposition} \label{prop:worst-case} Define for  generic loss variables $\bm{L} =(L^1,\dots, L^J )$  such that $L^j$ takes values in the interval $[0,1]$ and   $\EV{L^j} = \bar \ell^j$  and fixed weights
$w^1,\dots,w^J$ summing to one the portfolio loss by $L = \sum_{j=1}^J w^j L^j$. Then it holds for $\kappa \in [0,1]$ that
\begin{equation}
E\Big( 1-L^* - \big( \kappa - L^* \big)^+\Big) \le E \Big( 1-L - \big( \kappa -L \big)^+ \Big ).
\end{equation}
\end{proposition}
\noindent The proof can be found in Appendix~\ref{app:bounds}.

We now discuss several economic implications.  First, under the worst-case default scenario the  LGD of all sovereigns is almost surely equal to one.   Note that under  constraints on the expected loss  a high LGD implies a low value for the default probability of  a given sovereign, so that $\bm{L}^*$ corresponds to a default scenario with `few but large losses'. Second, the worst-case default scenario maximizes the probability of large default ``clusters''  given the expected loss constraints. This is explained in detail in Appendix~\ref{app:bounds} where we discuss properties of the distribution ${\pi}^*$ of $\bm{L}^*$.  Third, note that it is possible to approximate the worst-case default scenario by properly parameterized versions of the   model introduced in Section~\ref{sec:model}; a precise construction is given in Appendix~\ref{app:bounds}. This shows that  it is possible to generate  arbitrarily conservative valuations for ESBies in our setup.

The qualitative properties of  $\bm{L}^*$  suggest that, in the dynamic default model from Section~\ref{sec:model}, an ESB is more risky for a given expected-loss level of the sovereigns if one  chooses  high values for  the mean reversion level of the default intensities  in the recession state $K$, so that many defaults are quite likely in that state;  at the same time the generator matrix has to be parameterized in such a way  that state  $K$  is visited relatively infrequently in order to meet the expected loss constraints.   This  intuition  underlies the construction  of the \emph{crisis scenarios} in the numerical experiments reported in the next sections. More generally, Proposition~\ref{prop:worst-case} gives a theoretical justification for the qualitative properties of ESBie prices observed in Section~\ref{sec:staticRisk} and in the work of \citeasnoun{bib:barucci-brigo-19}, \citeasnoun{bib:brunnermeier-et-al-17} or  \citeasnoun{bib:ESRB-18}: for a given expected-loss level of the sovereigns higher default correlations and a higher LGD leads to a higher expected loss for ESBies.

\subsection{Expected Loss of ESBies}\label{sec:staticRisk}

From now on we consider  ESBies with a time to maturity of five years and, for simplicity, a risk free  interest rate $r=0$.
In order to make the prices of ESBies with different attachment points $\kappa$ comparable, we consider normalized  ESBies with payoff
$\frac{1}{1-\kappa} \min( V_T, 1-\kappa)$, so that  the payoff of a normalized ESB is equal to one if there is no default, i.e.~for $L_T  \le \kappa$.
Moreover, we  introduce the risk-neutral \emph{expected tranche loss}
\begin{equation}\label{eq:def-exp-tranche-loss}
 \ell^{\ESB,\kappa}(0, X_0, \bm{\gamma}_0, \bm{L}_0 \, ; \, Q, \bm{\mu}) = 1- \frac{1}{1-\kappa}h^{\ESB,\kappa}(0, X_0, \bm{\gamma}_0, \bm{L}_0\, ; \, Q, \bm{\mu})\,.
\end{equation}
Here $\bm{\mu} = \{\mu^j(k)),\, 1\le k \le 3, \,1 \le j\le J \} $,   $Q$ is the generator matrix of $X$ and the function  $h^{\ESB,\kappa}\left(t,X_t, \bm{\gamma}_t, \bm{L}_t \, ; \, Q, \bm{\mu}\right)$ gives the price of an ESB with attachment point $\kappa$ at time $t$, see equation~\eqref{eq:def-ESB-price}. We have made the parameters $Q$ and $\bm{\mu}$ explicit in \eqref{eq:def-exp-tranche-loss} since we want to study how variations in their values affect the expected loss of ESBies. Note that we may interpret the annualized expected loss $\frac{1}{T}\ell^{\ESB, \kappa}$  as \emph{credit spread} $c^{\ESB,\kappa}(0,T)$ of a normalized ESB with attachment point $\kappa$.  In fact, since $r=0$ and since  for $x$ close to one $\ln x \approx x-1$, it holds   that
$$ c^{\ESB,\kappa}(0,T) = \frac{-1}{T} \ln \Big (  \frac{1}{1-\kappa}h^{\ESB,\kappa}\Big )  \approx \frac{1}{T} \ell^{\ESB,\kappa} \,.$$

\paragraph{Parameters.} As before, we work with  $K=3$ states of $X$. We choose the portfolio weights $w^j$ according to the GDP proportions within the euro area; numerical values are given in Table~\ref{tab:exposure} in Appendix~\ref{app:calibration}.
We use the mean LGD from Table~\ref{tab:lgd}, the volatility parameters $\sigma^j$ and the calibrated  trajectories $\{\bm{\gamma}_{s_m}\}$ and $\{X_{s_m}\}$ obtained  in Section~\ref{sec:calibration}.  In our numerical experiments  we consider three different parameter sets and the worst-case distribution $\pi^*$ (the distribution of the worst case default scenario from Proposition~\ref{prop:worst-case}). In the \emph{base parameter set}  we use the generator matrix from  Section~\ref{sec:calibration}.  We take $\omega^j =0$\footnote{Using a different value for $\omega^j$ has only a very minor impact on the spread and the loss probability of ESBies.} and calibrate  $\bm{\mu}$ and $\kappa^j$ to the full CDS term structure at the valuation date, so that the parameterized model  accurately reflects  the market's expectation at that date.\footnote{The calibration in Section~\ref{sec:calibration}, on the other hand, yields  a fixed set of  parameters giving  a reasonable fit throughout the entire observation period. This   provides evidence for the good performance  of our model in explaining market data, but is of course subject to small pricing errors at any given date.}

The generator matrix $Q$ is hard to calibrate from historical data, essentially since products depending on the default correlation of euro area countries   are not  traded. To deal with the ensuing model risk, we introduce two  \emph{crisis parameter sets}. In  these
parameterizations  the recession state (state three) occurs less frequently than under  the base parametrization, but if it occurs default intensities are (on average) substantially larger than for the base parameter set.   To achieve this, we consider two generator matrices $\widetilde{Q}_1$ and $\widetilde{Q}_2$  chosen  such  that, on average, $X$ spends less time in  state three than under the base parametrization.   The corresponding mean reversion levels $\widetilde{\bm{\mu}}_1$ and $\widetilde{\bm{\mu}}_2$ and   are determined from the constraint that  the expected loss $E(L_T^j)$ is identical  for all parameter sets; this typically leads to  $\mu^j(3) < \widetilde{\mu}^j_1(3) < \widetilde{\mu}^j_2(3)$.
The  entries of $\widetilde{Q}^1$,  $\widetilde{Q}^2$ are  provided in Table~\ref{tab:Q2Q3}.

\begin{figure}[h]
   \centering
\includegraphics[width=0.9\textwidth]{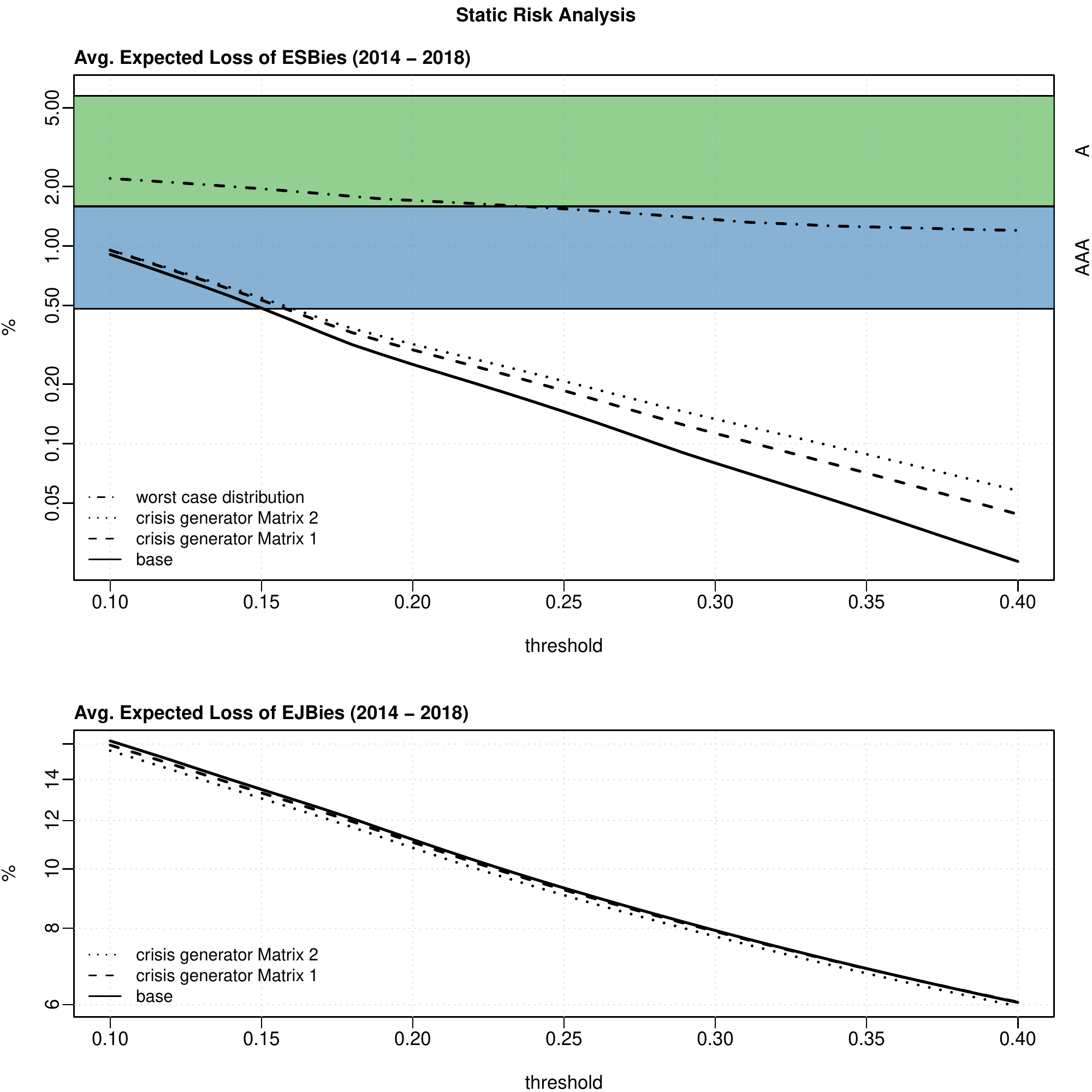}
    \caption{\small \label{fig:staticRisk}Average expected loss of ESBies (top) and of EJBs (bottom) for different thresholds and parameterizations (in \%). Note that both graphs use a logarithmic scale on the $y$-axis.}
\end{figure}

\paragraph{Results.} In the top panel of Figure~\ref{fig:staticRisk}, we graph the average expected loss\footnote{%
Here the term ``average'' refers to the average over observation dates, but with a fixed time to maturity of five years, that is we plot the function  $\kappa \mapsto \frac{1}{M} \sum_{m=1}^M \ell^{\ESB,\kappa}(0, X_{s_m}, \bm{\gamma}_{s_m}, \bm{0} \, ; \, Q, \bm{\mu})$.}
of ESBies over the period from 2014 to September 2018 as a function of the threshold $\kappa$. We do this for the base parametrization, the two crisis parameterizations and the worst-case distribution from Proposition~\ref{prop:worst-case}. The scale for the $y$-axis is logarithmic and  values are given in percentage points.
In addition, we consider AAA- and A- rated sovereigns (DEU, NLD and IRL, ESP, respectively) and compute the 1\%- and 99\%-quantile of the risk-neutral  expected loss over the period from 2014 to September 2018. Those quantiles form the boundaries of the colored areas in Figure~\ref{fig:staticRisk}; they  are supposed to give an indication of  the credit quality for the ESBies on a rating scale.\footnote{We stress that these indicative ratings should not be taken as actual ratings of ESBies, since they are computed from risk-neutral expected losses and not from historical ones, and since a rating is more than a mechanical mapping of expected loss to some rating scale.}

From Figure~\ref{fig:staticRisk} we draw the following conclusions. First, the average  risk-neutral expected loss of ESBies is indeed small. For example, the average expected loss corresponding to the proposed attachment point of 0.3 is below 0.1\%. Most strikingly, except for the worst-case distribution, the average expected loss of ESBies with thresholds of 0.15 or higher is well below the lower bound of the AAA-region.
Second, the expected loss is lowest for the base parameters, followed by crisis parameterizations~1 and 2; this  is fully in line with the economic intuition underlying the construction of these parameter sets.
Third, the expected loss for the  worst-case distribution (which is highest by construction)  is substantially higher than the expected loss in the crisis parameterizations, underlining the fact that the worst-case distribution, and the associated weak-link approach of  \citeasnoun{Kraemer2017HowESBies}, are extremely conservative.   Nonetheless, for $\kappa > 0.25$ the average expected loss for the worst-case distribution is still comparable in size to that of AAA-rated sovereigns.
Fourth, the expected loss of an ESB  is  decreasing  approximately at an  exponential rate in $\kappa$ in all four parameter sets (recall that  we use a logarithmic scale for the $y$-axis). Summarizing, these  findings show that an investor willing to hold  ESBies with an attachment point of 0.15 or higher until maturity faces little risk of default-induced losses, which is in agreement with  the analysis  of \citeasnoun{bib:brunnermeier-et-al-17} or \citeasnoun{bib:barucci-brigo-19}.

The bottom panel of Figure~\ref{fig:staticRisk} shows the average expected loss of EJBies for varying attachment points.
With five-year expected loss levels ranging from 6\% to  around  15\% (and hence annualized credit spreads between 1.2\% and 3\%)
the risk of EJBies is comparable to that of lower-quality euro area sovereigns.
Comparing the expected loss of ESBies and EJBies, we see that, in line with the proposal of \citeasnoun{bib:brunnermeier-et-al-17}, EJBies  bear the bulk of the credit risk associated to the eurozone sovereigns.
Note that the reverse ordering of the lines in the two panels of Figure~\ref{fig:staticRisk} is an immediate consequence of the put-call parity relation~\eqref{eq:put-call-parity}.

\subsection{Spread Trajectories of ESBies}\label{sec:ESB_ts}

\begin{figure}[h]
   \centering
\includegraphics[width=0.9\textwidth]{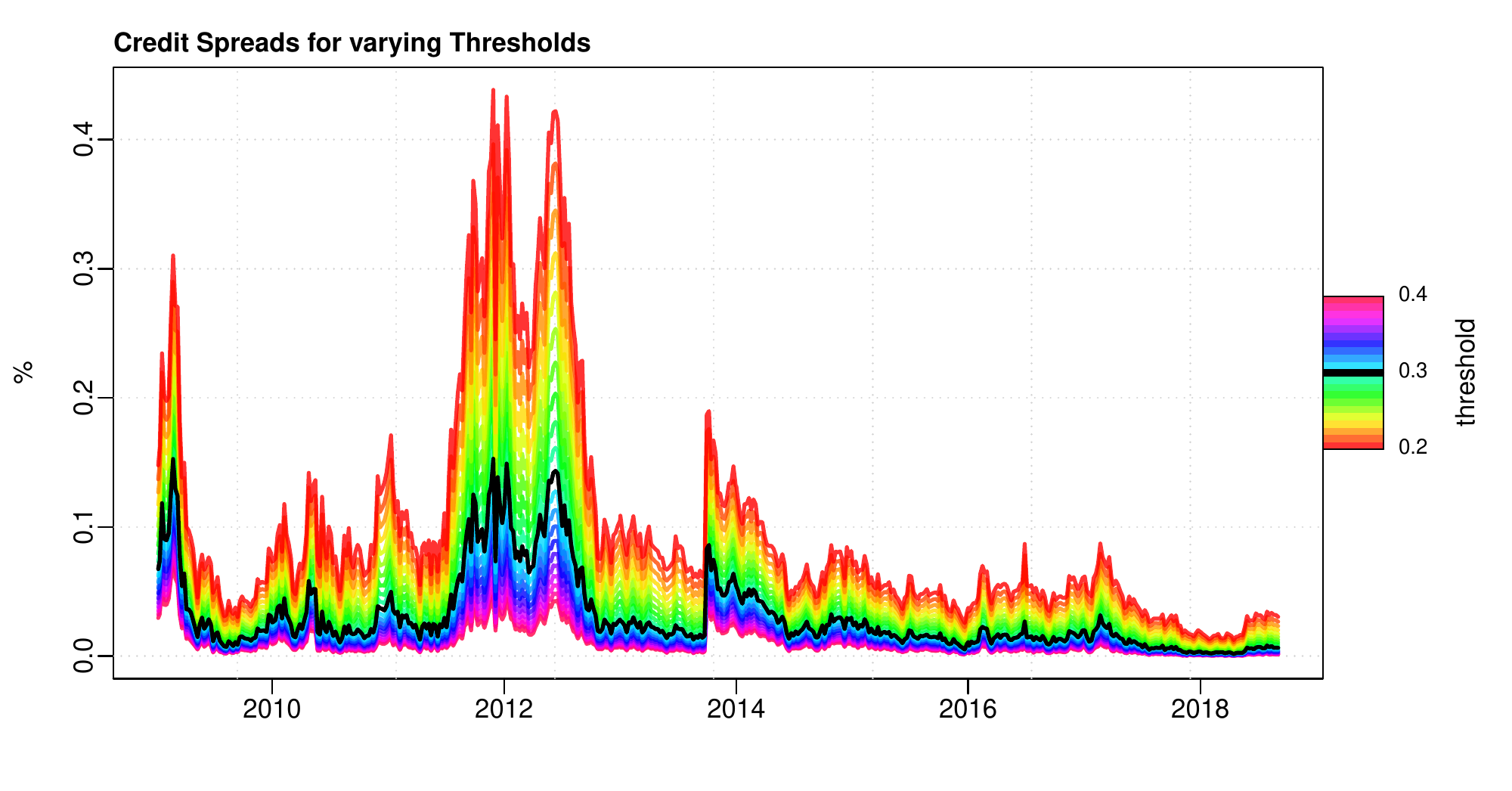}
    \caption{\small \label{fig:ESB_ts}{Spread trajectories of ESBies with varying threshold levels.} The solid black line represents the reference threshold of 0.3.}
\end{figure}

In Figure~\ref{fig:ESB_ts} we plot trajectories of the annualized credit spread $ c^{\ESB,\kappa}$ of  ESBies over the whole sample period for different levels of $\kappa$. These spreads were computed from our model by a historical simulation approach  using  the calibrated trajectories  $\{\bm{\gamma}_{s_m}\}$ and $\{X_{s_m}\}$ as input.
The solid line gives the spread of an ESB with attachment point $\kappa = 0.3$ (the value proposed by \citeasnoun{bib:brunnermeier-et-al-17}); the colored lines correspond to different attachment points $\kappa \in [0.2, 0.4]$.  The simulated ESB spreads peak in  2009 (the height of the financial crisis) and in the period 2011--2013 (the height  of the European sovereign debt crisis).   We see that the attachment point has a large impact on the volatility  of ESB spreads. In particular for $\kappa$ close to 0.2 spreads are very volatile; for $\kappa>0.3$ on the other hand spread fluctuations are quite small.

\subsection{Market risk analysis via scenarios}\label{sec:scenarios}

In this section, we  analyze how  the risk-neutral loss probability $\Q(L_T >\kappa)$ of ESBies is affected by changes in the underlying risk factors $X_0, \bm{\gamma}_0$ and $\bm{L}_0$. In mathematical terms, we consider the function
$$\kappa \mapsto p^\kappa(X_0, \bm{\gamma}_0, \bm{L}_0 \,;\, Q, \bm{\mu}) := \Q(L_{T} > \kappa \mid X_0, \bm{\gamma}_0, \bm{L}_0 \,;\, Q, \bm{\mu} ). $$
We consider different sets of risk factor changes or \emph{scenarios}. First we study scenarios  which are included in the support of the default model from Section~\ref{sec:model}, such as a change in the state of $X$.  Moreover,   we consider several \emph{contagion scenarios} where, in reaction to a default of Italy,\footnote{We consider a default of Italy since on the day we used for this analysis (September 3, 2018, the last observation date in our sample) Italy had the highest CDS spread of all major euro area economies. A default of another major `risky' euro area sovereign would yield similar results.}
the market becomes more risk averse and changes  its perception of the state of $X$ and the parameter set used to value ESBies. In fact,  investors on markets for (quasi) safe assets frequently  change their expectations in reaction  to adverse events, putting more mass on bad outcomes; see for instance~\citeasnoun{bib:gennaioli-shleifer-vishny-12}.

\begin{figure}[h]
    \centering
\includegraphics[width=0.8\textwidth]{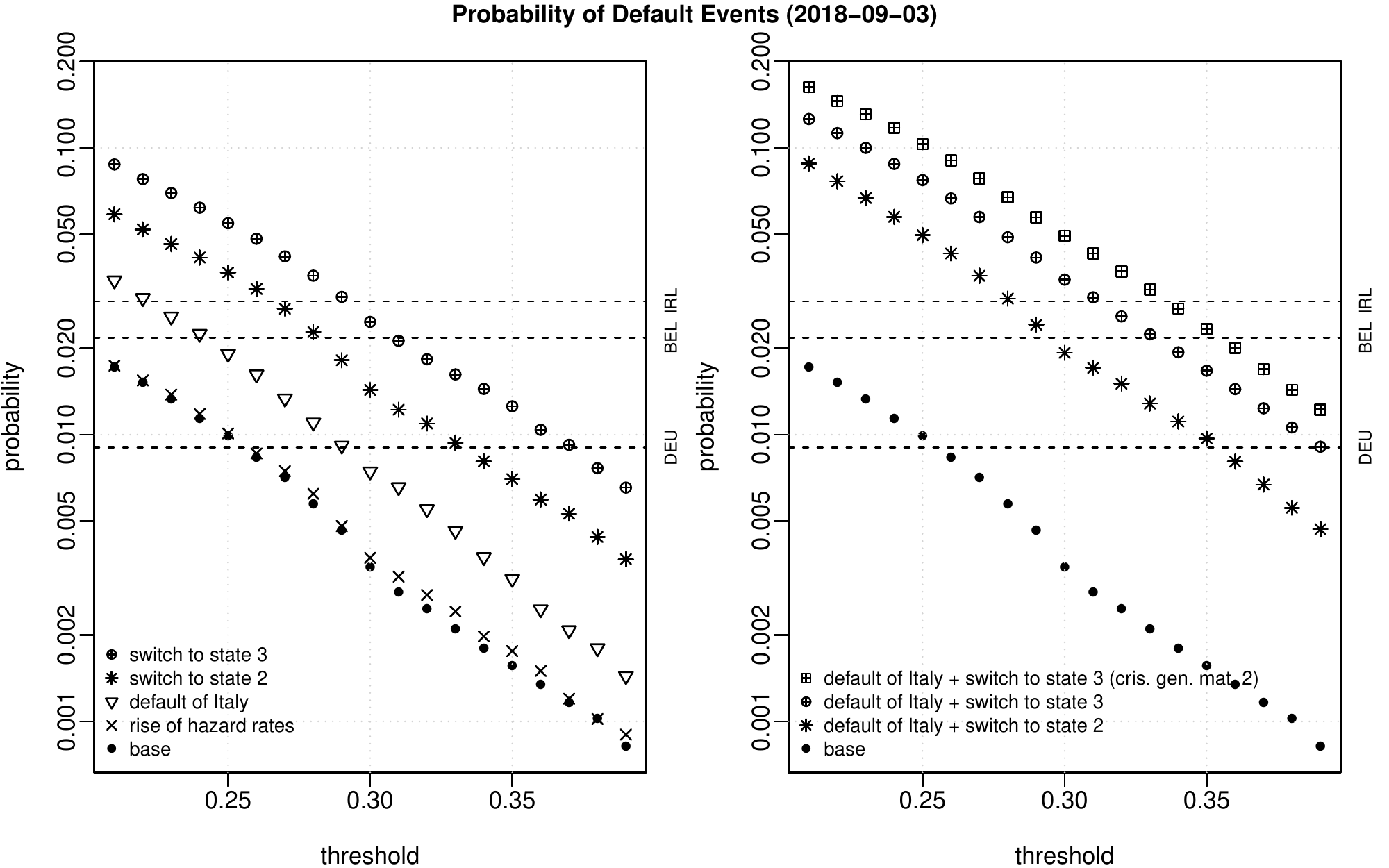}
    \caption{\small \label{fig:scenario}Loss probability of ESBies for different $\kappa$ and various scenarios. Note that the plot uses a logarithmic scale on the $y$-axis. }
\end{figure}

\paragraph{Non-contagion scenarios.}
In the left panel  of Figure~\ref{fig:scenario}, we graph the function $p^\kappa$ {on a log-scale} using  the parameters of the base scenarios and the calibrated values of $\bm{\gamma}$ and $X$ for September 3, 2018 (the last observation date in our sample). The full circles give the loss probability for varying $\kappa$ for the \textit{base scenario}, where the chain is in state one (the good economic state) and no euro area member is in default (in mathematical terms this is the function $\kappa \mapsto p^\kappa(1, \bm{\gamma}_0, \bm{L}_0 \,;\, Q, \bm{\mu})$).
Moreover, we consider four types of changes in the underlying risk factors:
\begin{itemize}
\item[(i)] the scenario  where all hazard rates experience an upward jump of 10\%, that is we plot the function $\kappa \mapsto p^\kappa(1, \bm{\gamma}_0 \times 1.1, \bm{L}_0 \,;\, Q, \bm{\mu})$;
\item[(ii)] the scenario where the economy moves to a light recession, corresponding to the function  $\kappa \mapsto p^\kappa(2, \bm{\gamma}_0, \bm{L}_0 \,;\, Q, \bm{\mu})$;
\item[(iii)]  the scenario  where the economy moves to a severe  recession ($\kappa \mapsto p^\kappa(3, \bm{\gamma}_0, \bm{L}_0 \,;\, Q, \bm{\mu})$);
\item[(iv)] the scenario where Italy defaults with random LGD $\delta^\text{ITA}$. We assume that $\delta^\text{ITA}$ is beta distributed with mean $0.5$, i.e.\ the loss vector at $t=0$ takes the form $\bm{L}_0 = (0, \dots, 0, \delta^\text{ITA}, 0, \dots, 0)$, but all other risk factors stay unchanged.
\end{itemize}
The horizontal dashed lines correspond to the risk-neutral five year default probabilities  of Germany, Belgium and Ireland under the  base parametrization.

Inspection of the left panel of Figure~\ref{fig:scenario} shows first that  a change in the hazard rates has only a small impact on the loss  probability of ESBies. The default of a major euro area sovereign such as Italy has a stronger effect, but for  $\kappa> 0.25$, the loss probability remains small even after a major default. 
The most important risk factor changes are clearly changes in the state of the economy. For instance, for $\kappa =0.3$ the loss probability of an ESB in state three is slightly larger than the risk-neutral default probability of Belgium, whereas in state one the loss probability is considerably smaller than the risk-neutral default probability of Germany.
Second, we observe that for the  given range of $\kappa$ the threshold probabilities are  decreasing in $\kappa$ roughly at an exponential rate, similarly as  the expected loss does.\footnote{We found this exponential decay for a wide range of parameter values,  but we do not have a fully convincing theoretical justification  for this effect.}
In fact,  the loss probability is quite sensitive with respect to the choice of the attachment point (to see this, one may compare the values of $p^\kappa$ for $\kappa =0.35$ and $\kappa= 0.3$ in scenario (iii)).

\paragraph{Contagion scenarios.} In the right panel  of Figure~\ref{fig:scenario} we graph the function $p^\kappa$ (again on a log-scale) for the base parametrization and for three different  contagion scenarios, namely
\begin{itemize}
\item[(i)] the case where Italy defaults and where, as a reaction,  $X$   jumps  to state two (mild recession);
\item[(ii)] the case where Italy defaults and where, as a reaction, $X$ jumps to state three (strong recession);
\item[(iii)] {the case where Italy defaults and where, as a reaction,  $X$ jumps to state three and the market uses the crisis parametrization two (instead of the base parameter set)}. This scenario is motivated by the observation that, in the subprime crisis, investors used much more conservative assumptions for default dependence than before the crisis, see for instance~\citeasnoun{bib:brigo-et-al-10} for details.
\end{itemize}
We see that, for an  attachment point $\kappa \le 0.3$, the change in the loss probability  caused by  one of the contagion scenarios is quite substantial. For instance, in the extreme scenario (iii), the risk-neutral loss probability is of the order of 5\%. For attachment points $\kappa >0.35$ the impact is less drastic. However, under scenario (iii), even for $\kappa = 0.35$ we  get a risk-neutral threshold probability of around 2\%, which is definitely non-negligible for a safe asset.
This is in stark contrast to the analysis of the expected loss in Section~\ref{sec:staticRisk}, where ESBies appeared `safe' already for $\kappa > 0.15$.


\subsection{Pooling  senior national tranches and robustness with respect to the LGD distribution} \label{subsec:tranche-and-pool}

\citeasnoun{bib:leandro-zettelmeyer-19} and a few other recent contributions suggest an alternative approach for constructing a safe asset for the euro area.  In these proposals,  the euro area sovereigns issue national bonds in (at least) two tranches, a senior and a junior tranche. A safe asset is then formed by pooling the senior tranche of the national debt, so that we use the acronym  PSNT (pooled senior national tranche) for these products.\footnote{From a financial engineering viewpoint also the E-Bonds of \citeasnoun{bib:monti-10}  fall in the category of national tranching followed by pooling; ESBies on the other hand  correspond to pooling followed by tranching.}
In this section we compare the risk-neutral expected loss and the risk-neutral loss probability of PSNTs to those of ESBies.
In particular, we focus on the impact of the random LGD since we cannot fully calibrate the distribution of the LGD from available market data.
We begin with a formal description of  the payoff of PSNTs. Given  some attachment level $\kappa$ that marks the split between the junior and the senior national bond tranches,  we model the payoff of the senior tranche issued by sovereign $j$ as
$
L_T^{j,\kappa} := (1-\kappa) - (L_T^j -\kappa)^+.
$
Note that  the senior national tranche suffers a loss  if $L_T^j$ exceeds the threshold $\kappa$.
For fixed  weights $w^1,\dots, w^J$, the  payoff of the PSNT  is then given by
$$
\sum_{j=1}^J w^j L_T^{j,\kappa} = (1-\kappa) - \sum_{j=1}^J w^j  (L_T^j -\kappa)^+  \,.
$$
Hence the  PSNT consists of a safe payment of size  $1-\kappa$  and a short position in a weighted  portfolio of  options  on the national losses. The normalized risk-neutral expected loss or equivalently the non-annualized  credit spread of a PSNT is given by
$$
\frac{1}{1-\kappa} \sum_{j=1}^J w^j E\big (( L_T^j -\kappa)^+ \big)\,.
$$
It follows that the credit spread of PSNTs is  independent of the dependence  structure of the national losses $L^1_T, \dots, L^J_T$.
Assumptions on the distribution of the loss given default, on the other hand, have a huge impact on the spread of PSNTs.  We begin with a few qualitative observations:  first,  it is easily seen that for fixed expected loss $E(L^j_T)$, the option price $E \big( (L_T^j -\kappa)^+\big)$   is maximal if $L_T^j \in \{0,1\}$. Hence, for fixed expected loss  level of the sovereigns, the expected loss of a PSNT is maximal if the LGD of all sovereigns is equal to one. In fact, in that  case the tranching on the national level offers no additional protection for the PSNT compared to simply pooling the  national bonds.
Second, if the LGD of all sovereigns is almost surely smaller than $\kappa$, the PSNT is entirely riskless. Finally, due to the convexity of the function $\ell \mapsto (\ell - \kappa)^+$ the expected loss of the PNST increases with increasing variance  of the LGD distribution (keeping $E(L_T^j)$ fixed).

\begin{figure}[h]
    \centering
\includegraphics[width=0.8\textwidth]{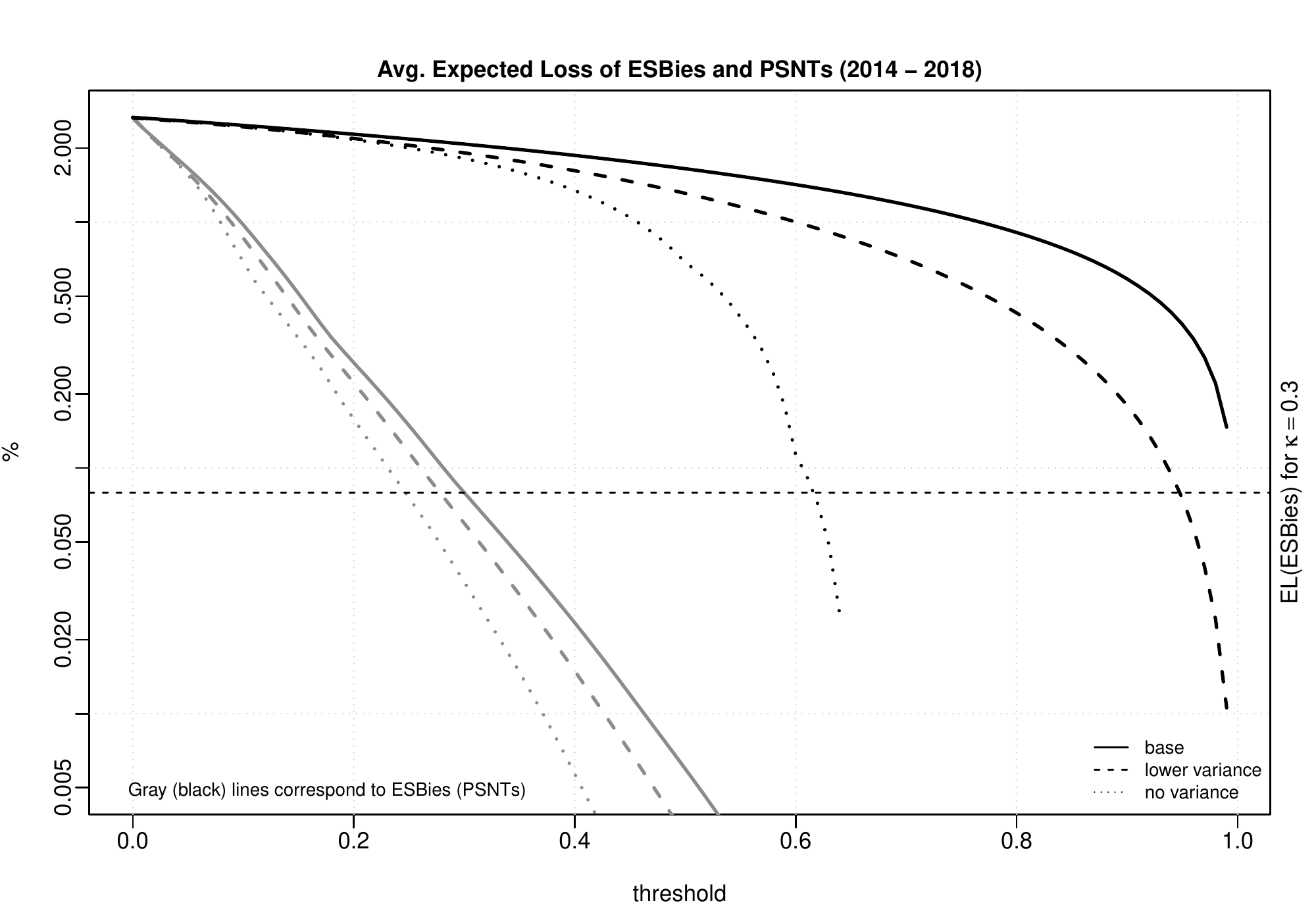}
    \caption{\small \label{fig:spread-ESB-PNST}  Spread of ESBies (grey) and of PSNTs (black) for varying  $\kappa$ and different value for the concentration parameter of the LGD distribution. The graph labelled base case corresponds to $\nu= 1.5$ (the same value as in Figure~\ref{fig:staticRisk}); the low variance case corresponds to the higher value $\nu=3.3$ of the concentration parameter.
    We fix $T = 5$. Note that the plot uses a logarithmic scale on the $y$-axis. }
\end{figure}

Next, we provide quantitative results comparing the behavior of PSNT credit spreads to that of ESBies.
Throughout this section, whenever we vary the mean of the random LGD, we also recalibrate the remaining model parameters such that any considered LGD setup is still in line with market data.
Figure~\ref{fig:spread-ESB-PNST} shows the average spread for ESBies (grey) and for PSNTs (black) over the period 2014-2018 for three different  LGD distributions with identical mean function given in Table~\ref{tab:lgd} and different variance/concentration parameter. We make the following observations. First,  the  spread of PNSTs is very sensitive to assumptions on  the variance of the LGD distribution whereas  the spread  of ESBies is comparatively stable. Second,  for the given mean function the  spread of PNSTs is substantially higher than for ESBies. In fact, even with deterministic LGD the expected loss of an ESBie with $\kappa=0.3$ equals the expected loss of a PNST with $\kappa \approx 0.6$; with higher LGD variance the two expected losses are equal only if the attachment point of the PSNT is close to one. The difference between the spreads of  ESBies and of PSNTs is due to  the fact that the  default of a single  euro area sovereign is sufficient to cause a loss for a PSNT, whereas ESBies are only affected in a severe default scenario with multiple defaults.   Moreover, for the payoff of a  PSNT it makes a substantial difference  if a sovereign  defaults only on its junior bond tranche or on both tranches which explains the sensitivity with respect to the LGD variance.

\begin{figure}[h]
    \centering
\includegraphics[width=0.8\textwidth]{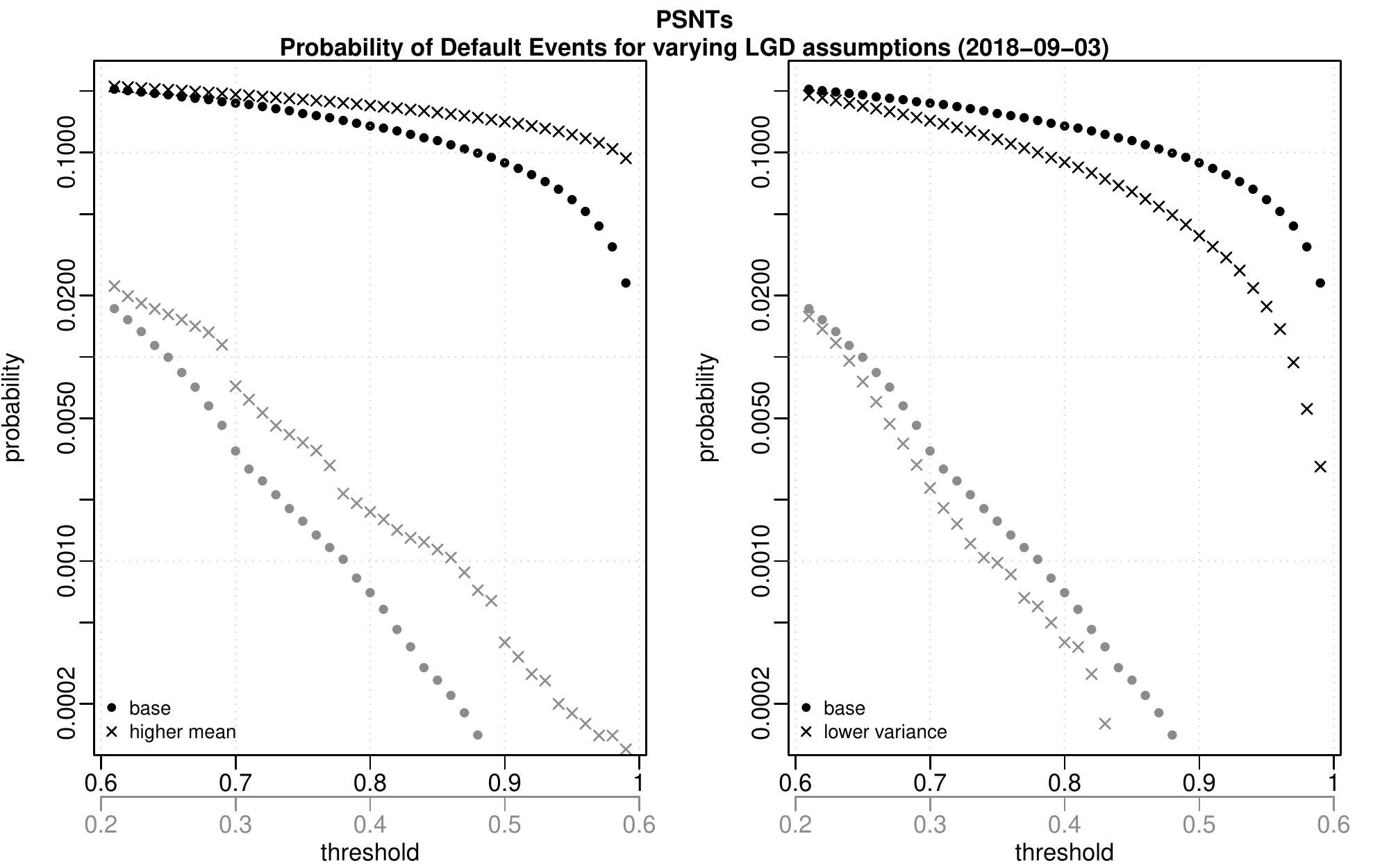}
    \caption{\small \label{fig:def-prob-PSNT} Loss probability of PSNTs (black) and ESBies (grey)  for varying  $\kappa$ and different value for the mean of the LGD distribution (left) and for the concentration parameter $\nu$ of the LGD distribution (right). The graph labelled base case corresponds to $\nu= 1.5$ and expected loss as in Table~\ref{tab:lgd}. For the higher mean setup, we increase the mean LGD of state~2 (state~3) by 0.2 (by 0.3) for all countries. The lower variance case uses the base mean and $\nu = 3.3$.
    We fix $T = 5$.
     The plot uses a logarithmic scale on the $y$-axis and different scales on the $x$ axis (grey for ESBies and black for PSNTs). }
\end{figure}

Finally, we consider the risk-neutral loss probability. Note first that  the risk-neutral loss probability of PSNTs is affected by the  dependence  structure of $L^1_T, \dots, L^J_T$ (other than the spread). In fact, for fixed marginal distributions of the sovereign losses, the  risk-neutral loss probability of PSNTs \emph{decreases} with increasing default correlation. This is akin to the behaviour of the equity tranche in standard CDO structures.
In Figure~\ref{fig:def-prob-PSNT} we  graph  the risk-neutral loss probability of ESBies and PSNTs for various values of the mean and the concentration parameter of the LGD distribution.  We observe the following:
first, for the parameter values considered the loss probability of PSNTs is substantially higher than that of ESBies. Second,  changes in the mean and in the concentration parameter have a profound impact on the loss probability of PSNTs; for ESBies these effects are less pronounced.
For a $\kappa = 0.3$, the loss probability of an ESB increases from $0.0043$ in the base setup to $0.0072$ for the higher mean case (which is still below the loss probability of a German bond, cf.\ Section~\ref{sec:scenarios}).
In comparison, for a relatively high $\kappa = 0.9$, the risk-neutral loss probability of an PSNT increases from $0.089$ to $0.142$.


\subsection{Market Risk Analysis via Loss Distributions }\label{sec:marketRisk}

So far we were  concerned with the \emph{value} of ESBies and EJBies in different scenarios; values were computed using the risk-neutral measure $\Q$, so that model parameters were  derived via calibration.
On the other hand, for computing measures of  market risk for the return of ESBies, we have to simulate their loss distribution  under the historical measure $\P$, so that we need to estimate the $\P$ dynamics of $X$ and $\bm{\gamma}$ using statistical methods.
This issue is addressed next.

\paragraph{EM estimation of hazard rate dynamics.} 
In this section, we report the results of  an empirical study where a model of the form \eqref{eq:basic-model} is estimated from  the  calibrated hazard rates of the  euro area countries (the trajectories $\{\gamma_{s_m}\}$ generated in the calibration procedure of Section~\ref{sec:calibration}). Here we assume that the trajectory of the Markov chain is not directly observable; rather, the available information is carried by the filtration $\mathbb{F}^{\bm{\gamma}} = (\mathcal{F}_t^{\bm{\gamma}})_{t \geq 0}$ generated by the hazard rates process $\bm{\gamma}$.  This assumption is motivated by the fact that the calibration of the trajectory $\{X_{s_m}\}$ in Section~\ref{sec:calibration} is quite  sensitive with respect to the chosen model parameters, whereas  the calibration of  $\{\bm{\gamma}_{s_m}\}$ is very robust (essentially due to the close connection between hazard rates and one-year CDS spreads).

Using stochastic filtering and a  version of the EM algorithm adapted to our setting, we obtain the filtered and  smoothed estimate for the trajectory of $X$, an estimate of the  generator matrix of $X$ and of  country-specific parameters such as mean reversion levels and speed, all under the real-world measure $\P$.    In the EM algorithm we use  \textit{robust} filtering techniques, which perform well in a situation where observations are only approximately of the form \eqref{eq:basic-model}. For further details on the methodology see \citeasnoun{bib:elliott-93} or \citeasnoun{bib:damian-eksi-frey-18}.

\begin{figure}
\begin{center}
\includegraphics[width=0.8\textwidth]{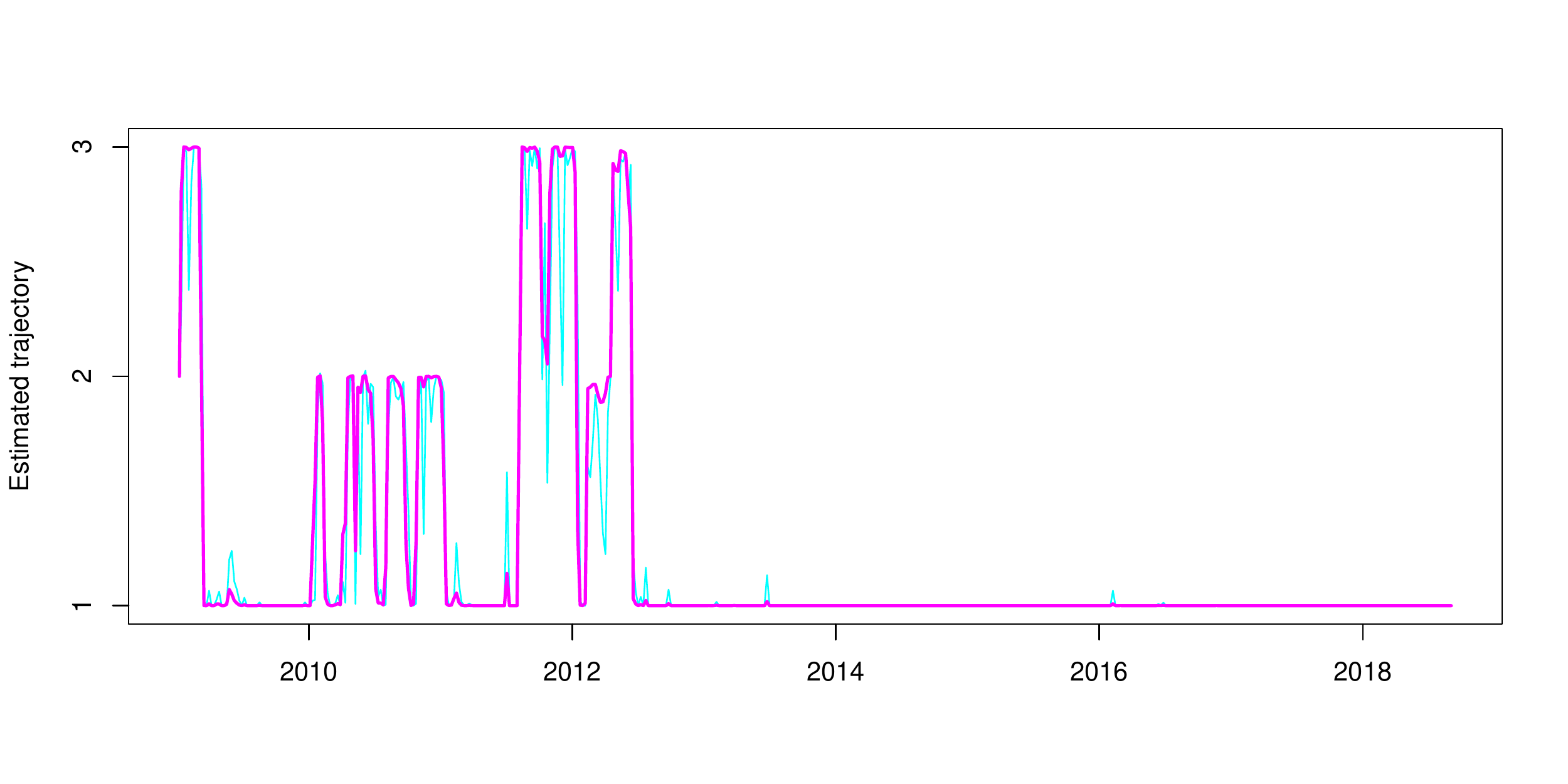}
\end{center}
\caption{\small \label{fig:EM} Filtered ($E^{\P}\left(X_t | \mathcal{F}_t^\gamma \right)$) and smoothed ($E^{\P}\left(X_t | \mathcal{F}_T^\gamma \right)$) estimates of the Markov chain trajectory.}
\end{figure}

We consider $K = 3$ possible states of $X$, corresponding to a expansionary regime,  a light recession and a strong recession, respectively. The EM estimates for the generator matrix $Q$ of $X$ are given in Tables \ref{tab:EMparam} and \ref{tab:EMQ} in Appendix~\ref{app:calibration}, together with country-specific parameters such as mean reversion speed and levels.
Note that we do not estimate the  volatility, but we work with  quadratic variation instead.\footnote{In order to robustify the EM estimation procedure, we scale the  quadratic variation of the strong euro area countries slightly upward.}
Overall the estimates appear reasonable. In particular, the estimated mean reversions levels for most countries respect the ordering $\mu^j(1) < \mu^j(2) < \mu^j(3)$, supporting  the interpretation of the states of $X$.\footnote{For Spain and Portugal, the highest mean reversion level is estimated for state~2, which is probably due to the idiosyncratic behavior these two countries, particularly Portugal, exhibit in the first months of 2012.} As expected, for any given state of the economy the estimated levels are lowest for the stronger euro area  countries.
In  Figure~\ref{fig:EM}, we give a trajectory of the filtered and the smoothed state of $X$, that is we plot the trajectories  $t \mapsto E^{\P}(X_t \mid \F_t^{\bm{\gamma}} )$ and   $t \mapsto  E^{\P}(X_t \mid \F_T^{\bm{\gamma}} )$. These  results show that the proposed model describes the qualitative properties of euro area credit spreads and, in particular, the co-movement of spread levels of the weaker euro-area members reasonably well. The frequent transitions in and out of  the middle state are not  surprising, given that  this state reflects a situation where only a few countries experience a rise in default intensities.

\paragraph{Measures of market risk.} We use two popular risk measures, Value at Risk ($\operatorname{VaR}_\alpha$) and Expected Shortfall ($\operatorname{ES}_\alpha$) at confidence level $\alpha$, to study the tail of the loss distribution of ESBies over a horizon of three months.
Denote by $\bm{\gamma}_0$ and $X_0$ the calibrated hazard rates and the state of $X$ for September~3, 2018. We generate $N = 100\,000$ realizations of the hazard rates and the Markov chain  with initial values $\bm{\gamma}_0$ and $X_0$ over a three-month horizon, using the  $\mathbb{P}$-parameters estimated in the previous paragraph, and we index the simulation outcome by $i \in\{1,\dots,N\}$.   We then compute the corresponding \emph{relative loss}
$${R}^{\kappa,i} := 1 - \frac{h^{\ESB, \kappa}(0.25, X_{0.25}^{(i)}, \bm{\gamma}_{0.25}^{(i)}, \bm{L}_0)}{h^{\ESB, \kappa}(0, X_0, \bm{\gamma}_0, \bm{L}_0)}.$$
VaR and expected shortfall are then computed from the empirical distribution of the sampled relative losses $\{{R}^{\kappa,i}, i=1,\dots,N\}$, see \citeasnoun[Section~9.2.6]{bib:mcneil-frey-embrechts-15} for details.

\begin{figure}[h]
    \centering
\includegraphics[width=0.85\textwidth]{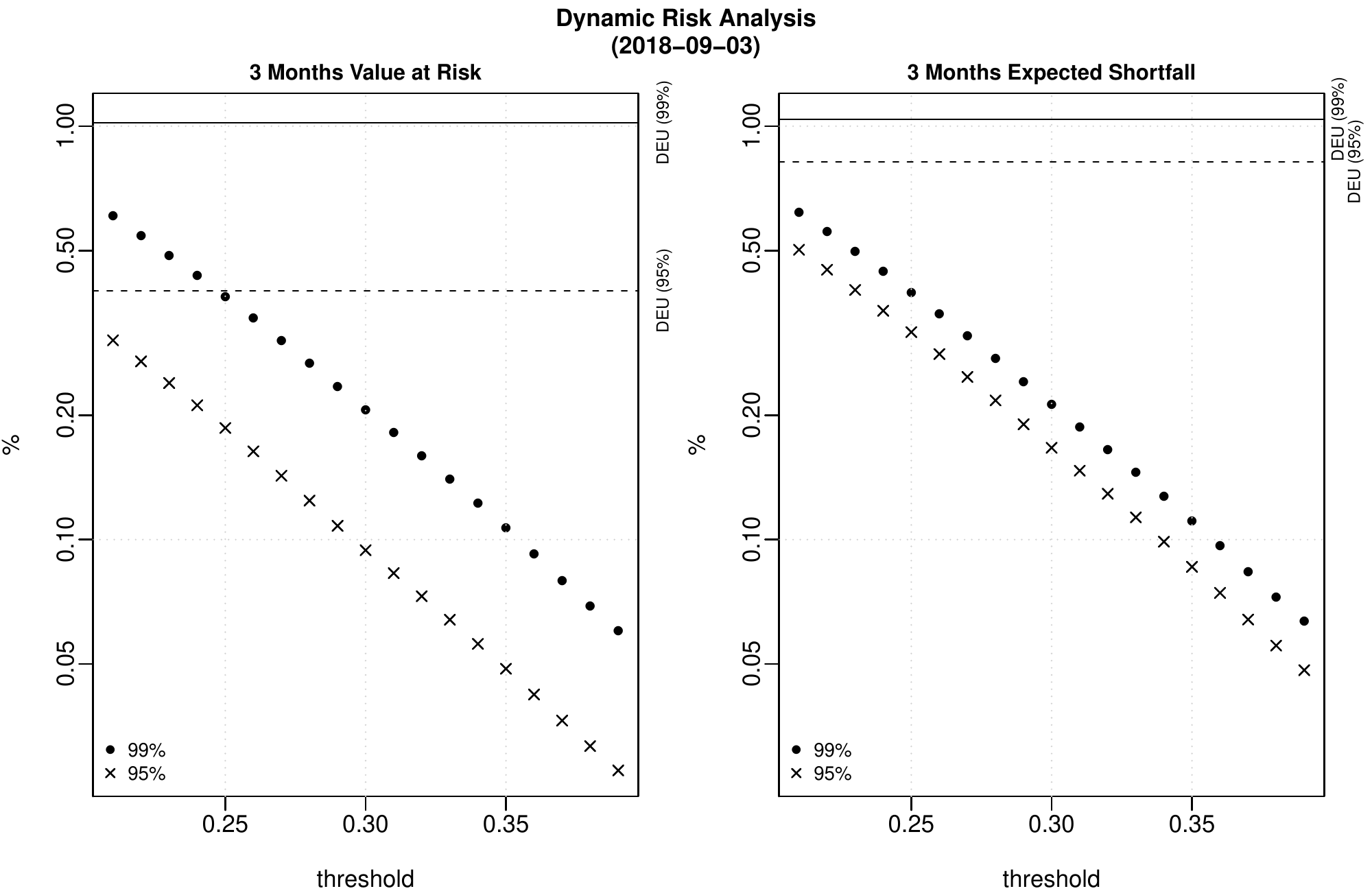}
    \caption{\small \label{fig:VaR_ES} Risk measure estimates $\operatorname{VaR}_\alpha$ (left) and $\operatorname{ES}_\alpha$  (right)  for the three-month  distribution  of negative ESB-returns for different $\kappa$ and confidence levels $\alpha \in \{0.95,0.99\}$.  Note that risk measures are given in percent and  that the plot uses a logarithmic scale on the $y$-axis. }
\end{figure}

Figure~\ref{fig:VaR_ES} summarizes our analysis. We plot  estimates of $\operatorname{VaR}_\alpha$ (left) and of $\operatorname{ES}_\alpha$  (right)  for the three-month  distribution  of negative ESB-returns for different $\kappa$ and confidence levels $\alpha =0.95$  (points) and $\alpha =0.99$ (crosses).
We see that both risk measure estimates  decrease approximately at an exponential rate in $\kappa$.
The horizontal lines in each plot represent the 95\% and 99\% level of the corresponding  risk measure estimate for  a German zero coupon  bond.
We observe that both the $\operatorname{VaR}_\alpha$ and the $\operatorname{ES}_\alpha$ of ESBies with $\kappa \geq 0.2$ are considerably smaller than the German benchmark.
We conclude that, from the viewpoint of a standard market risk analysis, ESBies appear safe and that very low risk capital levels are required to back these products. This observation is fully in line with the market risk analysis of \citeasnoun{bib:de-sola-perea-et-al-19}, who  find that ``(spread changes for) senior tranches have low tail risk exposure, often lower than the tail risk of lowest-risk euro area sovereigns.'' This coincidence is interesting since the analysis of \citeasnoun{bib:de-sola-perea-et-al-19} uses a  completely different methodology, namely a time-series approach based on GARCH models and a VAR for VaR analysis.

\section{Summary and Policy Implications}\label{sec:summary}
We draw the following  key conclusions from the risk analysis of ESBies and PSNTs.
Both the static risk analysis of Section~\ref{sec:staticRisk} and the investigation of the loss distribution in Section~\ref{sec:marketRisk} suggest that, in normal circumstances, diversification works and ESBies with $\kappa >0.25$ are indeed very  safe products.\footnote{In fact, from the perspective of an expected loss analysis, already an attachment point $\kappa =0.15$ might suffice to make ESBies safe.} In line with this finding, we showed that the weak link approach for the rating of ESBies proposed by S\&P is extremely conservative as it corresponds to a worst-case default scenario. We have also seen that for typical parameter values, the   spread and the loss probability of ESBies   are substantially lower than those of  PSNTs (securities created by pooling the senior tranche of national debt).  Moreover, for PSNTs, spread and loss probability are more sensitive to changes in the LGD distribution than for ESBies. This shows that from a risk perspective ESBies might be preferable to PSNTs.

However,  considering solely the results of Section~\ref{sec:staticRisk} and Section~\ref{sec:marketRisk} could lead to an overly optimistic picture. The analysis of credit spread trajectories in Section~\ref{sec:ESB_ts} and the  scenario based analysis of Section~\ref{sec:scenarios} show  that  the attachment point $\kappa$ needs to be chosen more conservatively  in order to make ESBies robust with respect to  fluctuations in the underlying risk factors or to changes in the market perception of default dependence. In fact, one has to take  attachment points $\kappa >0.35$ for ESBies to be safe {even in very adverse scenarios}. Moreover, ESBies are most likely to generate large market  losses in the aftermath of severe economic shocks and in  contagion scenarios.

From a policy perspective, it is therefore important that a large-scale  introduction of  ESBies is accompanied by  appropriate policy measures to  limit the economic implications of external shocks and  default events in the  euro area (and thus default contagion). {Such measures include a substantial weakening of  the  sovereign-bank nexus;   a completion of the banking  and capital markets union and the creation of a European deposit insurance scheme to improve risk sharing; more flexible forms of ESM (European Stability Mechanism) lending to countries in financial difficulties and bond clauses to allow for an orderly restructuring of sovereign  debt and a bail-in of private investors, and perhaps even limited direct transfers to countries hit by severe economic shocks, see \citeasnoun{bib:cepr-on-euro-18} for details.} In conjunction with  these measures,  the introduction of ESBies  would  be a useful step in improving the financial architecture  of the euro area.

\appendix

\section{Pricing methodology}\label{app:pricing}

Our main  tool for computing prices of credit derivatives  is the following extended Laplace transform for Markov modulated CIR processes. A related result was derived in \citeasnoun{bib:elliott-siu-09} for the case  a single CIR-type process, see also \citeasnoun{bib:vanBeek-et-al-20}
\begin{proposition} \label{prop:affine-transformation}
Denote by $\mathbb{F} = (\F_t)_{t \ge 0}$ the filtration generated by the Brownian motion $\bm{W}$ and the Markov chain  $X$.
Consider vectors $\bm{a}, \bm{u} \in \mathbb{R}_{+}^{J} $ and  a function  $\xi: S^X \to \mathbb{R}$.  Fix some horizon date $s \le T$.
Then it holds that for $0 \leq t < s$
\begin{equation}
\label{eqn:transform}
\EV{ \xi(X_s) \exp\left(- \int_t^s  \bm{a}' \bm{\gamma}_\theta  d\theta \right) e^{- \bm{u}' \bm{{\gamma}}_s } \mid \F_t} = v(t, X_t) \exp\left(\bm{\beta}(s-t, \bm{u})' \bm{{\gamma}}_t \right).
\end{equation}
Here $\bm{\beta}(\cdot, \bm{u}) = (\beta_{1}(\cdot, \bm{u}), \dots, \beta_{J}(\cdot, \bm{u}))'$ and the functions $\beta_{j}(\cdot, \bm{u}), \, 1 \leq j \leq J$, solve the Riccati equation
\begin{align}
    \label{eqn:beta}
   \partial_t \beta_{j}(t,\bm{u}) &= - \kappa^j \beta_j(t,\bm{u}) + \frac{1}{2}(\sigma^j)^2 \beta_j^2(t,\bm{u}) - a_j, 0 < t \le s \,,
\end{align} 
with initial condition  $\bm{\beta}(0, \bm{u}) = -\bm{u}$.  Moreover, with $\bm{v}(t) = \big(  v(t,1),\dots,v(t,K)\big)'$, the function $v \colon [0,s] \times S^X \rightarrow \mathbb{R}$ satisfies the linear ODE system
\begin{equation}
    \label{eqn:ODE}
  - \frac{d}{dt}\bm{v}(t) -\operatorname{diag} \left( \bar\mu_1(t), \dots, \bar\mu_K(t) \right)\bm{v}(t) = Q \bm{v}(t), \quad \text{on } [0,s],
\end{equation}
with terminal condition $\bm{v}(s) =  \bm{\xi}$ and with
$\bar\mu_k(t) = \sum_{j=1}^{J}e^{\omega^j t} \kappa^j\mu^j(k)\beta_j(s-t,\bm{u})$.
\end{proposition}
The functions $\beta_{j}(t,\bm{u})$ are known explicitly, see for instance \citeasnoun{filipovic2009term} for details. Essentially, Proposition~\ref{prop:affine-transformation} shows that computing the extended Laplace transform of $\bm{\gamma}$ is not much more complicated than in the classical case of independent CIR processes; the only additional step is to solve the  $K$-dimensional linear ODE system \eqref{eqn:ODE} for the function $\bm{v}(t)$, which is straightforward to do   numerically.

\begin{proof}
We start by conditioning in \eqref{eqn:transform} on $\F_t \lor \F_\infty^X$.
Due to the independence of the Brownian motions $W^1,\dots, W^J$, we have conditional independence of $\gamma^1, \dots, \gamma^J$ given $\F^X_\infty$, which in turn leads to
\begin{align}
    &\EV{ \xi(X_s) \exp\left(- \int_t^s  \bm{a}' \bm{\gamma}_\theta  d\theta \right) e^{- \bm{u}' \bm{{\gamma}}_s } \mid \F_t  \lor \F_\infty^X} \notag\\
    \label{eqn:proof_affine_aux1}
    &\qquad = \xi(X_s) \prod_{j = 1}^J \EV{\exp\left(- \int_t^s  a_j \gamma^j_\theta  d\theta \right) e^{- u_j {\gamma}^j_s } \mid \F_t  \lor \F_\infty^X}.
\end{align}
Conditional on $\F^X_\infty$, the hazard rates $\gamma^j$ are time-inhomogeneous affine diffusions.
Standard references on affine models, such as \citeasnoun{bib:duffie-pan-singleton-00}, consequently give that
\begin{equation}
    \label{eqn:proof_affine_aux2}
    \EV{\exp\left(- \int_t^s  a_j \gamma^j_\theta  d\theta \right) e^{- u_j {\gamma}^j_s } \mid \F_t  \lor \F_\infty^X} = \exp\left(\alpha_j(t, s \, ; \, X) + \beta_j(s - t, \bm{u}) \gamma_t^j \right),
\end{equation}
where $\beta_j$ solves \eqref{eqn:beta} and where $\frac{d}{dt}\alpha_j(t,s\,;X) = -e^{\omega^j t}\kappa^j \mu^j(X_t) \beta_j(s -t)$ and $\alpha_j(s,s\,;X) =0$;  see for instance \citeasnoun{bib:duffie-pan-singleton-00} or Section~10.6 of \citeasnoun{bib:mcneil-frey-embrechts-15} for a proof.
Integration thus gives
$ \alpha_j(t,s\,;X) = \int_t^s e^{\omega^j \theta} \kappa^j \mu^j(X_\theta) \beta_j(s - \theta) d\theta$.
By iterated conditional expectation, we hence  get
\begin{align*}
& \EV{ \xi(X_s) \exp\left(- \int_t^s  \bm{a}' \bm{\gamma}_\theta  d\theta \right) e^{- \bm{u}' \bm{{\gamma}}_s } \mid \F_t} \\
&\qquad =  \exp\Big (\sum_{j=1}^{J} \beta_j(s-t) \gamma_t^j\Big)
    \EV{\xi(X_s) \exp\left(\int_t^s \bar{\mu}_{X_\theta}(\theta) d \theta \right) \mid \F_t}
\end{align*}
The Feynman Kac formula for functions of the Markov chain $X$ finally gives that
$$
\EV{\xi(X_s) \exp\left(\int_t^s \bar{\mu}_{X_\theta}(\theta) d \theta \right) \mid \F_t} = v(t, X_t),
$$
and hence the result.
\end{proof}

Next we consider the pricing of a survival claim and of a CDS on sovereign $j$.

\paragraph{Survival claim.} The payoff of a survival claim on sovereign $j$  with maturity date $s$ and payoff function $f \colon S^X \to \R$   is of the form $\ind{\tau^j>s} f(X_s)$.
Using standard results on doubly stochastic default times, the price of this claim at time $t \le s$ is
$$ \EV{ B_{t, s}^{-1} \ind{\tau^j > s } f(X_s) \mid \G_t } =  \ind{\tau^j >t } B_{t, s}^{-1} \EV{  e^{-\int_{t}^{s}\gamma_{s}^j ds } f(X_s) \mid \F_t},$$
and the expectation on the right  can be computed from Proposition~\ref{prop:affine-transformation} with $\bm{a} = \bm{e}^j$, $\bm{u} = \mathbf{0}$ and ${\xi}  = {f}$.

\paragraph{Credit default swap.}
We briefly discuss CDS  pricing in our setup, since this is crucial for model calibration. From the payoff description~\eqref{eq:payoffCDS},
pricing a  CDS contract amounts to computing the conditional expectation
\begin{equation}
    \label{eqn:CDSexpect}
    \EV{\sum_{n = 1}^N B_{t, t_n}^{-1} \ind{\tau_j \in (t_{n-1}, t_n]} \delta^j_{t_n} - \sum_{n = 1}^N x (t_n - t_{n-1}) B_{t, t_n}^{-1} \ind{\tau_j > t_n} \mid \G_t}.
\end{equation}
Denote by  $V_t^\text{prem}$ and $V_t^\text{def}$ the present value of the premium and the default leg, that is
\begin{align}
       V_t^\text{prem}(x) &= \sum_{n = 1}^N B_{t, t_n}^{-1} x (t_n - t_{n-1}) \EV{\ind{\tau_j > t_n} \mid \G_t}  \notag , \\
        V_t^\text{def} &= \sum_{n = 1}^N B_{t, t_n}^{-1} \EV{\ind{\tau_j \in (t_{n-1}, t_n]}  \delta^j_{t_n}) \mid \G_t } \notag \\
                \label{eqn:defLeg}
        &=  \sum_{n = 1}^N B_{t, t_n}^{-1} \EV{\ind{\tau_j \in (t_{n-1}, t_n]}  \delta^j(X_{t_n}) \mid \G_t}\,.
\end{align}
To obtain \eqref{eqn:defLeg}, we have used the fact that the default leg of the CDS is linear in the loss given default, so that
we can replace  $\delta^j_{t_n}$ with its conditional expectation.
The premium leg is simply the sum of survival claims. The  evaluation of \eqref{eqn:defLeg} is more involved, and we now show how
 this can be achieved via Proposition \ref{prop:affine-transformation}.
Fix any two consecutive payment dates $t_{n-1}, t_n$ of $\mathbb{T}$  and assume w.l.o.g.~that  $t\le t_{n-1}$.
Since $\ind{t_{n-1} < \tau^j \leq t_n} = \ind{\tau^j > t_{n-1}} - \ind{\tau^j > t_n}$, we can write the term  $\EV{\ind{\tau_j \in (t_{n-1}, t_n]}  \delta_j(X_{t_n})\mid \G_t}$ in the form
\begin{equation}
    \label{eqn:defLeg_aux}
    \EV{\ind{\tau^j > t_{n-1}} \delta^j(X_{t_n}) \mid \G_t} - \EV{ \ind{\tau^j > t_n} \delta^j(X_{t_n}) \mid \G_t}.
\end{equation}
The second term in~\eqref{eqn:defLeg_aux} is a survival claim. By iterated conditional expectations, we get that the first  term is equal to
\begin{equation}
    \label{eqn:defLeg_aux2}
    \EV{\ind{\tau^j > t_{n-1}} \EV{\delta^j(X_{t_n}) \mid \G_{t_{n-1}}} \mid \G_t}.
\end{equation}
Since $X$ is Markov, it  holds that $\EV{\delta^j(X_{t_n}) \mid \G_{t_{n-1}}} = v^{\delta}(t_{n-1}, X_{t_{n-1}})$ for a suitable function $v^\delta \colon [0,t_n] \times S^X \to \R$ (given by the solution of an ODE system), so \eqref{eqn:defLeg_aux2} reduces to computing $  \EV{\ind{\tau^j > t_{n-1}} v^\delta(t_{n-1}, X_{t_{n-1}})\mid \G_t}$, which is a standard pricing problem for a survival claim.

Finally, we turn to the pricing of ESBies. In order to evaluate the function $ h^{\EJB,\kappa}$ we use Monte Carlo simulation.
For the computation of the function  $ h^{\ESB,\kappa}$ we use that $h^{\ESB,\kappa} = \EV{B_{t,T}^{-1} (1-L_T) \mid \G_t} - h^{\EJB,\kappa}$ and we compute the  expected discounted portfolio loss analytically.

\section{Worst-case default scenario and price bounds} \label{app:bounds}

In this section we provide some additional results underpinning   our discussion of the worst-case default scenario and  lower price bounds for  ESBies in Section~\ref{sec:price-range}.

\paragraph{Proof of \protect{Proposition~\ref{prop:worst-case}}.}
By the put call parity \eqref{eq:put-call-parity} for ESBies and EJBies, the claim of the proposition is equivalent to showing that  $\bm{L}^*$
maximizes the value of EJBies. More precisely, we show that
for  any random vector  $(L^1, \dots, L^J)'  \in [0,1]^J $ with $E(L^j) = \bar \ell^j$, $1 \le j \le J$, and any $\kappa >0 $, it holds that \begin{equation} \label{eq:minimization-problem}
E\Big ( \big ( \sum_{j=1}^J w^j L^j - \kappa \big)^+ \Big ) \le  E\Big ( \big (  \sum_{j=1}^J w^j L_j^* - \kappa \big)^+ \Big ).
\end{equation}
We may use call options instead of put options in \eqref{eq:minimization-problem} since $E(\sum_{j=1}^J w^j L^j)$ is fixed. To establish the inequality \eqref{eq:minimization-problem} we use  a result on stochastic orders  from \citeasnoun{bib:bauerle-muller-06}. According to the equivalence ((iii) $\Leftrightarrow$ (iv)) in  Theorem~2.2 of that paper, \eqref{eq:minimization-problem}  is equivalent to
the inequality
\begin{equation} \label{eq:minimization-problem-2}
\ES_\alpha\Big (\sum_{i=1}^J w^j L^j\Big)  \le \ES_\alpha\Big(\sum_{i=1}^J w^j L_j^*\Big) \text{ for all } \alpha \in [0,1)\,,
\end{equation}
where for a generic random variable $Z$,  $\ES_\alpha (Z) = \frac{1}{1- \alpha} \int_\alpha^1 q_u (Z) \,du$ gives the expected shortfall of $Z$ at confidence level $\alpha$ and where $q_u(Z)$ denotes the quantile of $Z$ at level $u$.

To establish \eqref{eq:minimization-problem-2} we show first that $L_j^* $ maximizes the quantity
$\ES_\alpha (L^j) $  over all rvs $L^j$ with value in the interval $[0,1]$   and expectation  $ E(L^j)= \bar \ell^j$, simultaneously for all $\alpha \in [0,1)$. In fact,  the random variable $L^j$ has to satisfy the constraints $q_u(L^j) \le 1$ (since $L^j \in [0,1]$) and $\int_0^1 q_u(L^j) \, du  =  \bar \ell^j$ (since $E(L^j) = \bar \ell^j$, so that
$$
\ES_\alpha(L^j)  \le \frac{1}{1- \alpha} \min \{ 1-\alpha,  \bar \ell^j\} = \ES_\alpha(L_j^*) \,.
$$
Moreover, we get from the coherence  of expected shortfall that
$$
\ES_\alpha\Big(\sum_{j=1}^J w^j L^j\Big) \le \sum_{j=1}^J w^j \ES_\alpha\big(L^j\big) \le \sum_{j=1}^J w^j \ES_\alpha\big(L_j^*\big) = \ES_\alpha\Big( \sum_{j=1}^J w^j L_j^*\Big)\,,
$$
where the last equality follows since  $L_1^*,\dots, L_m^*$ are comonotonic.  This gives inequality \eqref{eq:minimization-problem-2} and hence the result. \qed

\paragraph{Distribution of $\bm{L}^*$.} Next we discuss properties of the distribution $\pi^*$ of the worst-case default scenario.   This distribution is a discrete probability measure on $[0,1]^m$ which charges $J+1$ points; it is given by
\begin{align*}
\pi^*\big((1,\dots,1)\big) &= \bar{\ell}^1, \, \pi^*\big((0,1,\dots,1)\big) = \bar{\ell}^2 - \bar{\ell}^1,    \cdots, \,\pi^*\big((0,\dots,0,1)\big) = \bar{\ell}^J - \bar{\ell}^{J-1},\\ \pi^*\big((0, \dots,0)\big) &= 1 - \bar{\ell}^J\,.
\end{align*}
We call  $\pi^*$ the worst-case distribution. Note that, under $\pi^*$, the probability of large default ``clusters'' is maximal given the expected loss constraints. First, under $\pi^*$  the event  where all sovereigns default has probability  $\bar{\ell}^1$.  Since
$$ \Q( L^1 = \dots= L^J = 1) \le \Q(L^1 = 1) \le E (L^1) = \bar{\ell}^1\,,$$
this is the maximum value possible. Next, under $\pi^*$ the default scenario   where all sovereigns  except the first default has probability $\bar{\ell}^2 - \bar{\ell}^1$. It is easily seen that this  is the maximum possible value given the expected-loss constraints and the probability attributed to the first cluster (the cluster where all sovereigns default).  Similarly,   the probability of the ($n+1$)-th cluster, where all but the first  $n$  sovereigns default, is  maximal given the probability attributed to the first $n$ clusters.

Finally we sketch an approach for the approximation of the worst-case distribution $\pi^*$ within our model.
 Note first that, for $\kappa^j$ large and $\sigma^j$ small, the hazard-rate trajectory $(\gamma^j_t)_{0\let\le T}$ is essentially determined by the trajectory of $X$ and by the choice of the mean reversion level $\mu^j(\cdot)$, so that we concentrate on these quantities.   We consider a model with $K= J+1$ states of $X$ that correspond to the different default ``clusters" under $\pi^*$.  Choose  some large $n$ and define the mean reversion level $\mu^j(\cdot)$ by
$\mu^1 (1) =\dots=\mu^J(1) = \frac{1}{n}$;  $\mu^1(2)= \dots=\mu^{J-1}(2)=\frac{1}{n},\;  \mu^{J}(2) = n$; \dots;
$\mu^1 (J+1) =\dots=\mu^J(J+1) = n$. Note that in state $k$ the default probability of obligor  $1$ to obligor $J-k+1$ is  small, the default probability of obligor $J +2 -k$ up to obligor $J$ is large; that is, the state corresponds to the  $(J +2 -k)$-th default cluster.

Next we define the generator matrix of $X$. We assume that states 2 to $J+1$  are absorbing, so that $q_{ik}= 0$ for  $2 \le i \le J+1$ and all $k$.
Define probabilities $p_1,\dots, p_{J+1}$ by
$ p_1 = 1- \bar{l}^J$, $p_k = \bar{\ell}^{J+2-k} - \bar{\ell}^{J+1 -k}$ for  $2 \le k \le J $,  and finally $p_{J+1} = \bar{\ell}^1$, that is $p_k$ corresponds to the probability of the $(J +2 -k)$-th default cluster under $\pi^*$.
Since states $2,\dots,J+1$ are absorbing, we get for any valid choice for the first row of $Q$  that $\Q(X_T = 1) = e^{q_{11}T} $ and
$$ \Q( X_T =k) = (1 - e^{q_{11}T}) \frac{q_{1k}}{-q_{11}}, \quad k=2, \dots, J+1\,,$$
(recall $q_{11} = - \sum_{k=2}^{J+1} q_{1k}$). We want to choose $q_{12},\dots,q_{1J+1}$  so that  $\Q(X_T = k ) = p_k$ for all $k$. This gives
\begin{equation} \label{eq:Q}
 q_{11} = \frac{1}{T} \ln p_1  \, \text{ and } q_{1k}= -p_k \frac{q_{11}}{1-p_1}\,, k=2,\dots,J+1\,.
 \end{equation}
Since  $\sum_{k=1}^{J+1} p_k =1$, we get that $q_{11} = - \sum_{k=2}^{K+1} q_{1k}$ so that \eqref{eq:Q} defines indeed a valid generator matrix.
Moreover, for  $n \to \infty$, $\kappa^j \to \infty$ and $\sigma^j \to 0$,
$$\Q\big ( \ind{\tau^1\le T} = \dots=  \ind{\tau^{J-k+1} \le T}  = 0, \ind{\tau^{J-k+2} \le T} =\dots = \ind{\tau^{J} \le T} =1 \big)$$
converges to  $\Q(X_T =k) = p_k$ which gives the result by definition of the $p_k$.

\section{Details on Calibration} \label{app:calibration}

\subsection{Data}

In Table~\ref{tab:summaryCDS} below we present summary statistics of the data we use in the model calibration.
\begin{table}[ht]
\centering
\resizebox{\columnwidth}{!}{%
\begin{tabular}{@{\extracolsep{5pt}}lSSSSSSSSSS}

 {Yrs.} & {AUT} & {BEL} & {DEU} & {ESP} & {FIN} & {FRA} & {IRL} & {ITA} & {NLD} & {PRT} \\
 & {AA} &  {AA} &  {AAA} &  {A} & {AA} & {AA} & {A} &  {BBB}  & {AAA}  &  {BBB} \\
  \hline
& \multicolumn{10}{c}{Panel A: Mean} \\
1 & 31.071 & 44.063 & 12.819 & 113.637 & 13.273 & 26.590 & 204.380 & 115.934 & 20.269 & 307.199 \\
  2 & 38.341 & 54.593 & 16.700 & 138.474 & 17.780 & 35.106 & 220.752 & 143.971 & 25.232 & 346.413 \\
  3 & 45.016 & 66.843 & 21.918 & 153.769 & 22.148 & 44.961 & 224.791 & 165.881 & 30.413 & 352.693 \\
  4 & 54.657 & 77.339 & 29.151 & 165.480 & 28.219 & 56.684 & 223.055 & 181.786 & 38.267 & 354.438 \\
  5 & 61.675 & 85.437 & 34.562 & 174.373 & 33.048 & 65.906 & 222.409 & 193.090 & 43.902 & 359.586 \\
& \multicolumn{10}{c}{Panel B: Standard Deviation} \\
1 & 38.589 & 59.330 & 12.418 & 113.398 & 12.932 & 30.303 & 309.597 & 110.190 & 21.720 & 436.057 \\
  2 & 42.347 & 67.134 & 14.520 & 129.415 & 14.257 & 34.529 & 308.774 & 113.689 & 23.782 & 449.052 \\
  3 & 44.800 & 74.987 & 17.157 & 131.520 & 15.065 & 39.920 & 293.590 & 115.705 & 24.563 & 399.150 \\
  4 & 48.974 & 76.066 & 21.330 & 130.483 & 16.769 & 45.769 & 264.432 & 113.967 & 27.595 & 352.837 \\
  5 & 51.023 & 76.099 & 24.114 & 129.470 & 17.407 & 49.149 & 244.132 & 112.625 & 29.439 & 324.968 \\
  & \multicolumn{10}{c}{Panel C: Minimum} \\
  1 & 4.080 & 3.840 & 2.920 & 10.450 & 2.250 & 3.550 & 7.830 & 21.620 & 3.120 & 12.480 \\
  2 & 6.190 & 7.020 & 3.980 & 18.900 & 3.800 & 6.570 & 12.330 & 33.880 & 4.840 & 28.670 \\
  3 & 7.820 & 9.430 & 6.230 & 25.050 & 6.040 & 9.620 & 15.860 & 48.990 & 7.230 & 40.650 \\
  4 & 9.890 & 11.910 & 8.330 & 30.730 & 10.290 & 12.630 & 19.670 & 56.490 & 9.140 & 47.110 \\
  5 & 13.270 & 16.480 & 9.510 & 37.230 & 13.020 & 17.400 & 23.970 & 59.830 & 11.240 & 47.230 \\
    & \multicolumn{10}{c}{Panel D: Maximum} \\
  1 & 259.960 & 301.620 & 74.840 & 489.430 & 66.530 & 160.660 & 1629.340 & 619.540 & 110.870 & 2598.930 \\
  2 & 267.440 & 337.600 & 81.080 & 608.330 & 74.600 & 177.440 & 1614.480 & 591.030 & 125.040 & 2494.690 \\
  3 & 269.490 & 375.700 & 90.350 & 619.920 & 82.550 & 198.200 & 1572.800 & 581.050 & 130.650 & 2102.190 \\
  4 & 271.430 & 379.090 & 108.250 & 622.220 & 90.460 & 222.510 & 1419.750 & 575.930 & 132.710 & 1846.700 \\
  5 & 272.180 & 380.940 & 119.060 & 624.290 & 95.000 & 237.300 & 1318.590 & 573.030 & 136.960 & 1802.360 \\
   \hline
\end{tabular}
}
\caption{\small Summary statistics of CDS spreads (in bp).\label{tab:summaryCDS}}
\end{table}

\subsection{Methodology}
In order to determine the parameters $(\Theta^j, \sigma^j)$, $1 \le j \le J$, the generator matrix $Q$  and  the realised trajectories  $\{\bm{\gamma}_{s_m}\}$ and $\{X_{s_m}\}$, we use an iterative approach which is compactly summarized in Algorithm~\ref{alg:CalCDS} below.
We set $\bm{\Theta} = (\Theta^1, \dots, \Theta^J)$ and we use $\{\bm{\gamma}_{t_m}\}^{(i)}$, $\{X_{t_m}\}^{(i)}$ and $(\Theta^j)^{(i)}$ to denote the $i$-th estimate of the distinct variables within the iteration.

\vspace{0.1cm}

\begin{algorithm}[H]
\label{alg:CalCDS}
\DontPrintSemicolon
\SetAlgoLined
\KwData{Market CDS spreads for maturities $u \in \mathcal{T}$  for each sovereign $1 \leq j \leq J$}
\KwResult{Estimates for $\{\bm{\gamma}_{s_m}\}$, $\{X_{s_m}\}$ and $\bm{\Theta}$}
 Initialization for $\{\bm{\gamma}_{s_m}\}^{(0)}$, $\{X_{s_m}\}^{(0)}$, $(\bm{\Theta})^{(0)}$ and $Q^{(0)}$\;
 $i = 0$\;
 \While{$\sum_{j=1}^J \sum_{m=0}^M l^j\left((\gamma^j_{s_m})^{(i)}, ({\Theta}^j)^{(i)}, (\sigma^j)^{(i)}, Q^{(i)},{X}_{s_m}^{(i)}\right) \geq \epsilon$}{
 \For{$j\leftarrow 1$ \KwTo $J$}{
 \For{$m \leftarrow 0$ \KwTo $M$}{
   $(\gamma^j_{s_m})^{(i+1)} = \argmin_\gamma l_{s_m}^j(\gamma, ({\Theta}^j)^{(i)}, (\sigma^j)^{(i)}, Q^{(i)}, X_{s_m}^{(i)})$ \label{alg:optimGamma}\;
 }
  Estimate $(\sigma^j)^{(i+1)}$ based on the quadratic variation of $(\gamma^j)^{(i+1)}$\;
 }
 \For{$m \leftarrow 0$ \KwTo $M$}{
  ${X}^{(i + 1)}_{s_m} = \argmin_{x} \sum_{j=1}^J l_{s_m}^j\left((\gamma^j_{s_m})^{(i+1)}, ({\Theta}^j)^{(i)}, (\sigma^j)^{(i+1)},Q^{(i)}, x)\right)$ \label{alg:optimX}\;
 }
 Estimate $Q^{(i + 1)}$ via MLE based on ${X}^{(i+1)}$\;
\For{$j\leftarrow 1$ \KwTo $J$}{
  $({\Theta}^j)^{(i + 1)} = \argmin_{{\Theta}} \sum_{m=0}^M l_{s_m}^j\left((\gamma^j_{s_m})^{(i+1)}, {\Theta}, (\sigma^j)^{(i+1)}, Q^{(i+1)} {X}_{s_m}^{(i+1)})\right)$ \label{alg:optimParam}\;
 }
 Set $i \leftarrow i + 1$\;
 }
\caption{\small Detailed description of calibration step}
\end{algorithm}
\vspace{0.1cm}

The assumption of  conditionally independent defaults substantially facilitates the calibration procedure:  given  an estimate for $Q$ and $\{X_{s_m}\}$, estimation of $\{\gamma^j_{t_m}\}$  and of the parameter vector $\Theta^j$ can be done independently for each sovereign $j$.
We initiate the calibration by applying $k$-means clustering on the relevant CDS spreads to get an estimate for ${X}^{(0)}$.
For small maturities $T$, it holds that $\widehat{\cds}^j_{T} \approx \delta^j(X) \gamma^j$.
We use this approximation along with the initial estimate $X^{(0)}$ to get an estimate for $(\gamma^j)^{(0)}$ and we consequently solve the optimization problem of line~\ref{alg:optimParam} in Algorithm~\ref{alg:CalCDS} to obtain the initial value $\bm{\Theta}^{(0)}$.
To compute the  estimates for $\sigma^j$, we use that the quadratic variation of $\gamma^j$ satisfies
$$[\gamma^j, \gamma^j]_t = (\sigma^j)^2 \int_0^t \gamma^j_{s} ds, $$
and we approximate the integral with Riemann sums.
For a given (estimated) realisation of the Markov chain, we use the standard MLE estimator for continuous-time Markov chains  to get an estimate of $Q$.

The main numerical challenge in the application of   Algorithm~\ref{alg:CalCDS} is to solve the optimization problem
\begin{equation}
    \label{eqn:optim-param}
    \min_{{\Theta}} \sum_{m=0}^M l^j\left((\gamma^j_{s_m})^{(i+1)}, {\Theta}, (\sigma^j)^{(i+1)}, Q,{X}_{s_m}^{(i+1)})\right).
\end{equation}
We impose the restriction that all parameters are non-negative and, for regularization purposes, we set the lower bound of the mean-reversion speed $\kappa^j$ to $0.1$ for all $j$.
During the first iteration of Algorithm~\ref{alg:CalCDS}, we employ 
an algorithm for constrained optimization as presented in \citeasnoun{Runarsson2005SearchOptimization}.
The algorithm uses heuristics to escape local optima.
In order to refine the estimation, in the subsequent calibration steps (i.e.\ for steps $i > 1$) we use the local optimizer of \citeasnoun{Powell1994AInterpolation}, which provides a derivative-free optimization method based on linear approximations of the target function.
After successful convergence of Algorithm~\ref{alg:CalCDS}, we perform a final refinement step in which we keep all input variables except $\Theta^j$, $1 \le j \le J$, fixed.

%

\subsection{Results}

\begin{table}[h!]
\renewcommand*{\arraystretch}{1.2}
\centering
\resizebox{\columnwidth}{!}{%
\begin{tabular}{@{\extracolsep{5pt}}lSSSSSSSSSS}
 {State} & {AUT} & {BEL} & {DEU} & {ESP} & {FIN} & {FRA} & {IRL} & {ITA} & {NLD} & {PRT} \\
  \hline
1 & 0.55 & 0.55 & 0.50 & 0.55 & 0.50 & 0.50 & 0.55 & 0.50 & 0.50 & 0.55 \\
  2 & 0.55 & 0.55 & 0.50 & 0.55 & 0.50 & 0.50 & 0.55 & 0.50 & 0.50 & 0.55 \\
  3 & 0.65 & 0.65 & 0.60 & 0.65 & 0.60 & 0.60 & 0.65 & 0.60 & 0.60 & 0.65 \\
   \hline
\end{tabular}
}
\caption{\small Fixed conditional means of LGDs for different sovereigns and varying states.}
\label{tab:lgd}
\end{table}

The following figure illustrates the quality of the model fit for two different sovereigns.

\begin{figure}[h!]
\centering
\includegraphics[width=0.9\textwidth]{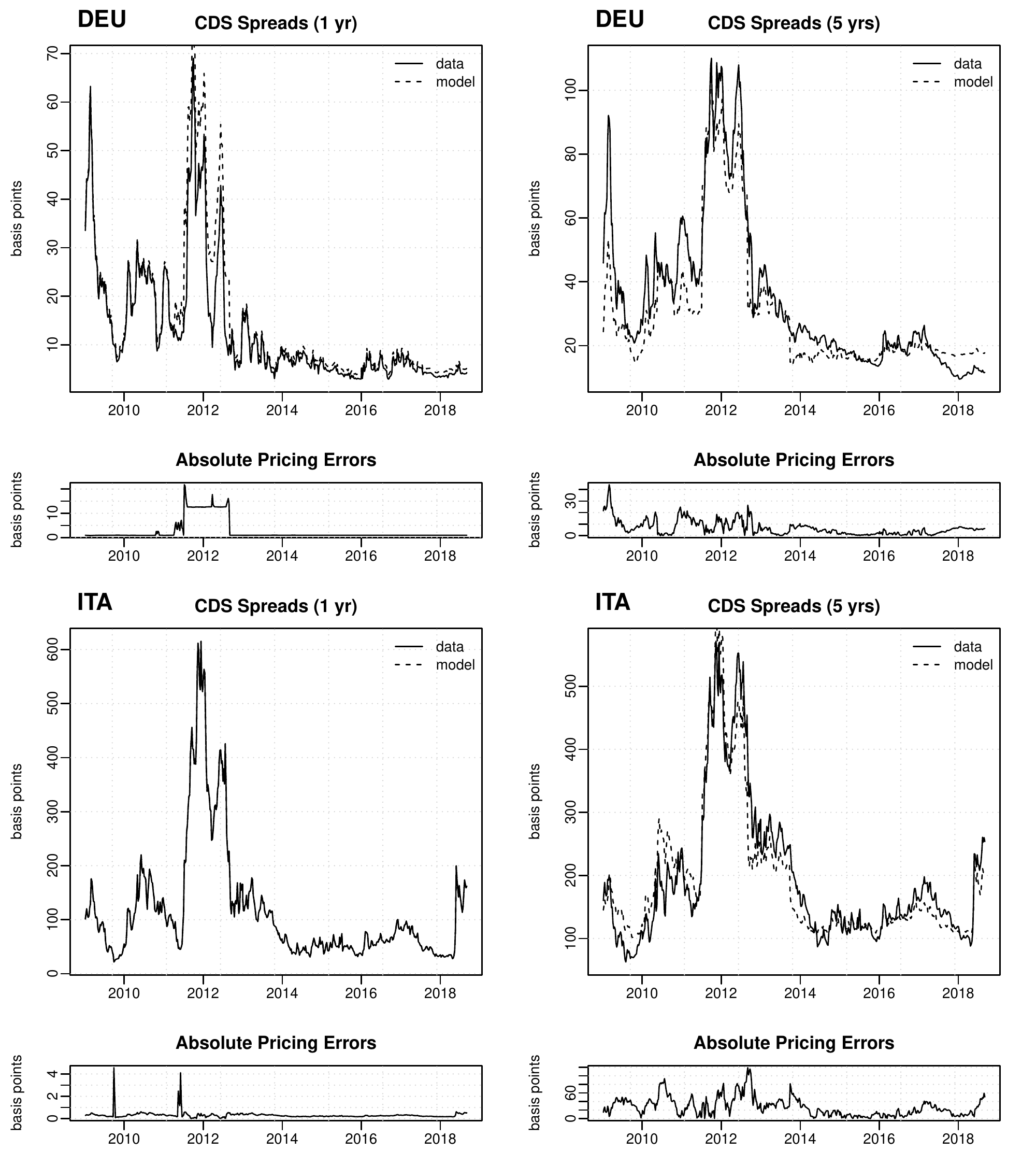}
       \caption{\small \label{fig:CDSseries} Time series plots of market CDS spreads against model values. The solid (dashed) lines correspond to the market (model) values of the distinct CDS spreads. }
\end{figure}

\subsection{Parameters used in Risk Analysis}

\begin{table}[h!]
\renewcommand*{\arraystretch}{1.2}
\centering
\resizebox{\columnwidth}{!}{%
\begin{tabular}{@{\extracolsep{5pt}}SSSSSSSSSS}
{AUT} & {BEL} & {DEU} & {ESP} & {FIN} & {FRA} & {IRL} & {ITA} & {NLD} & {PRT} \\
  \hline
0.04 & 0.04 & 0.29 & 0.12 & 0.02 & 0.20 & 0.03 & 0.18 & 0.07 & 0.01 \\
   \hline
\end{tabular}
}
\caption{\small Portfolio weights of ESBies and EJBies, based on proportion of sovereigns on euro area GDP as of 2018.}
\label{tab:exposure}
\end{table}

\begin{table}[h!]
\label{tab:Q2Q3}
\centering
\resizebox{\columnwidth}{!}{%
\begin{tabular}{l|SSS|SSS}
&\multicolumn{3}{c}{$\widetilde{Q}_1$} & \multicolumn{3}{c}{$\widetilde{Q}_2$} \\
\hline
  &{State 1} & {State 2} & {State 3} &{State 1} & {State 2} & {State 3} \\
   {State 1 (expansion)} & -0.1421 & 0.1421 & 0.0000  & -0.1421 & 0.1421 & 0.0000 \\
   {State 2 (mild recession)} &0.5843 & -0.8685 & 0.2843 & 0.5843 & -0.7685 & 0.1843 \\
   {State 3 (strong recession)} & 0.0000 & 1.4444 & -1.4444  & 0.0000 & 1.4444 & -1.4444 \\
   \hline
\end{tabular}
}
\caption{\small Generator matrices $\widetilde{Q}_1$ and $\widetilde{Q}_2$ for crisis scenarios.}
\end{table}

\subsection{Results of EM Estimation}
\begin{table}[h!]
\renewcommand*{\arraystretch}{1.2}
\centering
\resizebox{\columnwidth}{!}{%
\begin{tabular}{@{\extracolsep{5pt}}lSSSSSSSSSS}
 {Param.} & {AUT} & {BEL} & {DEU} & {ESP} & {FIN} & {FRA} & {IRL} & {ITA} & {NLD} & {PRT} \\
  \hline
  $\mu(1)$ & 0.0023 & 0.0016 & 0.0012 & 0.0013 & 0.0029 & 0.0220 & 0.0329 & 0.0095 & 0.0136 & 0.0027 \\
  $\mu(2)$ & 0.0103 & 0.0054 & 0.0013 & 0.0016 & 0.0196 & 0.1391 & 0.1375 & 0.0408 & 0.1231 & 0.0099 \\
  $\mu(3)$ & 0.0144 & 0.0120 & 0.0116 & 0.0080 & 0.0346 & 0.1219 & 0.0941 & 0.0698 & 0.1245 & 0.0192 \\
  \hline
  $\kappa$ &  6.9584 & 6.2427 & 3.1193 & 6.3241 & 5.5822 & 3.4718 & 1.5879 & 3.2518 & 2.0640 & 8.4694 \\
   \hline
\end{tabular}
}
\caption{ \small \label{tab:EMparam} Estimation results:  parameters of hazard rate dynamics.}
\end{table}

\begin{table}[h!]
\centering
\begin{tabular}{ lSSS}
                            &{State 1} & {State 2} & {State 3} \\
                            \hline
   {State 1 (expansion)} & -0.9033 &  0.9033 & 0.0000 \\
   {State 2 (mild recession)} & 5.9877 & -10.4716 & 4.4839 \\
   {State 3 (strong recession)} & 4.3316 &  1.8569 &-6.1885 \\
   \hline
\end{tabular}
\caption{\small \label{tab:EMQ} Estimation results:  generator matrix $Q$ of $X$.}
\end{table}


\end{document}